\documentclass[12pt,preprint]{aastex}

\shorttitle{pH and salt concentration in water planets' oceans}
\shortauthors{Levi, and Sasselov}

\usepackage{graphicx}
\usepackage{textcomp}
\usepackage{amssymb}
\usepackage{natbib}
\usepackage{amsmath}
\usepackage{subfigure}
\usepackage{multirow}\usepackage[section] {placeins}

\begin{document}
\pagenumbering{arabic}

\title{A New Desalination Pump Help Define the pH of Ocean Worlds}
\author{A. Levi}
\affil{Harvard-Smithsonian Center for Astrophysics, 60 Garden Street, Cambridge, MA 02138, USA}
\email{amitlevi.planetphys@gmail.com}
\author{D. Sasselov}
\affil{Harvard-Smithsonian Center for Astrophysics, 60 Garden Street, Cambridge, MA 02138, USA}

\maketitle

\section*{ABSTRACT}

We study ocean exoplanets, for which the global surface ocean is separated from the rocky interior by a high-pressure ice mantle.
We describe a mechanism that can pump salts out of the ocean, resulting in oceans of very low salinity. Here we focus on the H$_2$O-NaCl system, though we discuss the application of this pump to other salts as well. We find our ocean worlds to be acidic, with a pH in the range of $2-4$. We discuss and compare between the conditions found within our studied oceans and the conditions in which polyextremophiles were discovered. This work focuses on exoplanets in the super-Earth mass range ($2$M$_\Earth$), with water composing at least a few percent of their mass. Although, the principal of the desalination pump may extend beyond this mass range.

\section{INTRODUCTION}

The pH of an ocean is a fundamental index, needed for describing oceanic chemistry and biochemistry.
Exoplanets of various mass densities are being discovered. These range in composition from very poor to very rich in water. One may argue that Earth, being relatively poor in water, is an end member of this water planet family. Therefore, describing the pH evolution of Earth \citep[e.g.][]{Halevy2017} provides the framework for understanding the evolution of seawater pH on planets that are relatively poor in water.
In this work we consider the other end members in the water planet family, planets very rich in water. Understanding both end members is an important step toward a global framework for the pH of water planets.

\begin{figure}[ht]
\centering
\includegraphics[trim=0.2cm 12.0cm 0.1cm 2cm , scale=0.55, clip]{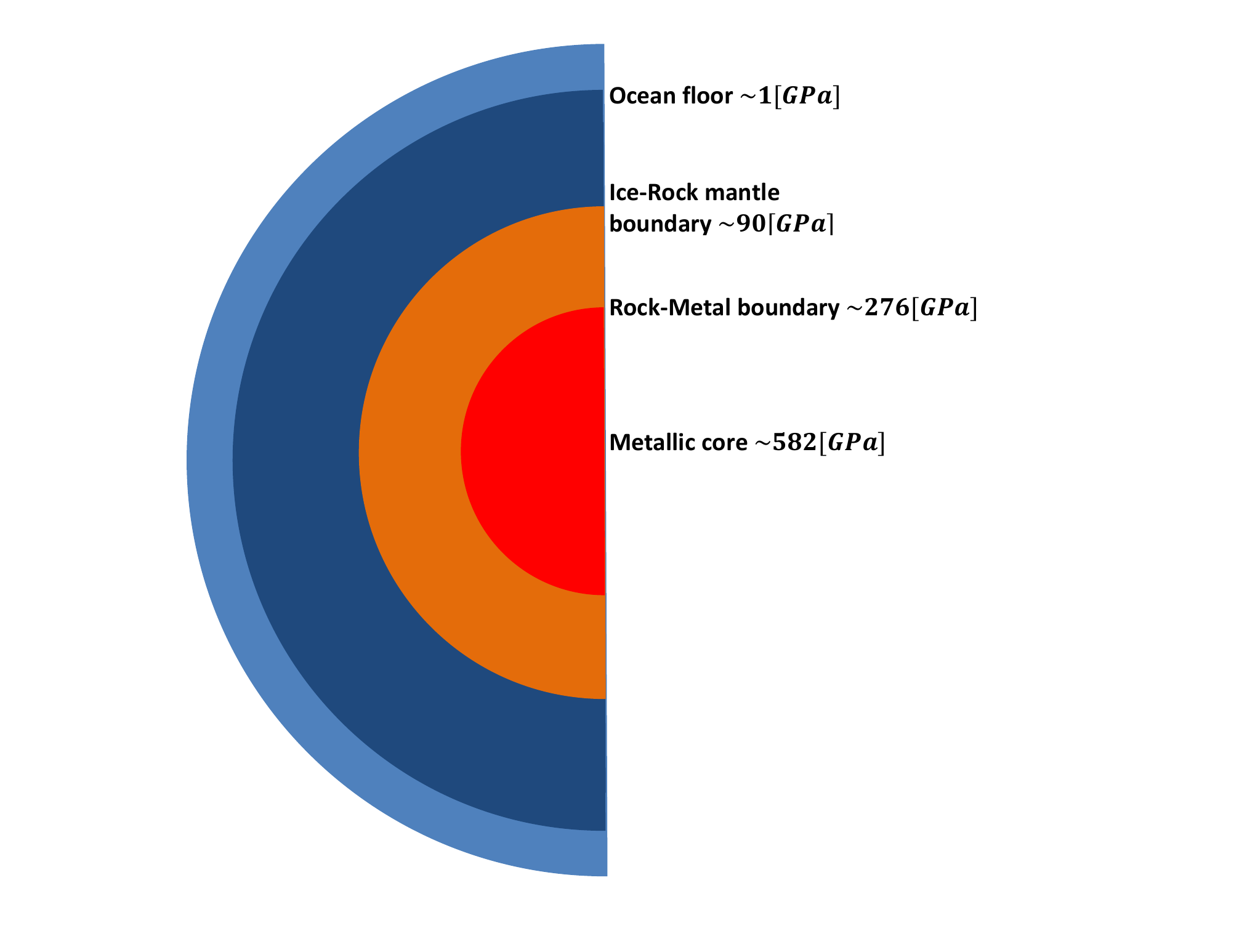}
\caption{\footnotesize{Cross section of a $2$M$_\Earth$ planet assuming $50$\% of its mass is H$_2$O. The rest of the mass is simplified to be $68$\% silicate perovskite inner rock mantle and $32$\% iron core. The inner part of the water mantle is a layer of high pressure ice, while the outer part is a global deep ocean.}}
\label{fig:CrossSection}
\end{figure}

We focus on exoplanets with a large mass fraction of water, and lacking a substantial primordially accreted H/He atmosphere. If the mass fraction of water is at least a few percent a deep high-pressure water ice layer forms on top of a rocky inner mantle \citep{Levi2014}. A representative illustration of the planets studied here is given in Fig.\ref{fig:CrossSection}. A high water mass fraction is a natural result of accretion beyond the snowline \citep{kuchner2003,leger2004}. The mass of the star and of the disk are correlated \citep{Andrews2013}. For Sun-like stars, which have relatively massive disks, accretion beyond the snowline results in a massive condensed core, capable of capturing H/He from the nebula. The high mass of the core prevents the photo-evaporation of such a primordial atmosphere. This is not the case if the condensed core is at most $2$M$_\Earth$ \citep{Luger2015}. However, accretion of a low mass condensed core beyond the snowline is more likely in a low mass disk. Therefore, the planets we study here are more likely to be found around M-dwarf stars \citep{Alibert2017}. The ubiquity of H$_2$O and of M-dwarfs suggest our studied planets should be very common. This range of masses is not well sampled with radial velocity measurements, it is however sampled by TTV's with a lower level of confidence. Examples for such planets may be Kepler $138b$ and $138d$ \citep{Jontof2015} and within the TRAPPIST-1 planetary system \citep{Gillon2017}. 

If these low mass ($<2$M$_\Earth$) water-rich planets migrate closer to their stars we expect their outermost condensed layer to be a global deep ocean. In addition their atmosphere is predominantly a secondary outgassed atmosphere. For such planets we wish to constrain their oceanic pH.            
CO$_2$ plays an important role in Earth's ocean. Being a weak acid it plays a dominant role in keeping the Earth's ocean neutral and sets the level of acidity, thus having major biological implications. We follow the same algorithm used for studying Earth's oceanic carbon speciation and acidity and apply it to oceans around water-rich exoplanets.

Earth's seawater is a complex solution of many ionic and neutral molecular species. The composition of seawater is due to an intricate system of sources and sinks, not all of them well understood. A major source for oceanic ions are the many rivers around the globe. These obtain their ions from: silicate weathering, dissolution of halite, gypsum and limestone, oxidation of organic matter and even from oceanic sprays recycling chlorine back inland and into rivers. A glimpse into the complexity of this system is revealed if one realizes that the ionic composition of rivers is much different from that of the ocean, the major components of the former are calcium and bicarbonate ions while of the latter are ions of sodium and chlorine. Therefore, one cannot obtain the composition of the open ocean simply by evaporating river aqueous solutions. Clearly, other sources and sinks ought be evoked in order to explain the composition of Earth's ocean. For example, atmospheric transportation of salts in aerosols from inland into the ocean seem to be of importance. Also, the ions produced in hydrothermal cycles are of considerable importance. In these cycles seawater penetrates the crust near mid-ocean ridges, and other active regions, through cracks forming in the basalt as it rapidly cools due to the direct contact with the descending seawater. The high temperatures in the upper crust ($300-400^\circ$C) yield many fluid-rock interactions and completely change the composition of the penetrating water. Eventually the high temperature increases buoyancy enough to force the water back up, where in some geographical locations the ejection of these waters takes the shape of black smokers, which forcefully discharge their altered composition into the ocean \citep[see chapter $13$ in][for an in depth discussion on the composition of Earth's ocean]{Pilson2013}. 

In the case of a super-Earth with a very small mass fraction of water ($<1\%$) a vast and deep ocean may indeed have a silicate bottom \citep{Levi2014}. In this case a few of the mechanisms just described for the case of Earth may be active. However, a planetary water mass fraction of a few percent, as in the objects of interest in this work, will suffice to create a high pressure ice barrier between any ocean and the rocky interior \citep{Levi2014}. In this case none of the above Earthly mechanisms will work.

The icy satellites of our solar system, for which a subterranean ocean is speculated, may also have a barrier made of high pressure water ice polymorphs separating the liquid layer from a differentiated/partly-differentiated rocky interior \citep[see][for Titan interior models]{Lunine2010book}. These satellites are considerably less massive than a water rich super-Earth, and therefore take much longer to differentiate. \cite{Lunine1987} estimated for the case of Titan that separation of ice from rock takes about $1$\,Gyr. Callisto probably still has regions of mixed ice and rock to this very day \citep{Anderson2001}. Even low temperature fluid-rock interactions can produce aqueous solutions very rich in magnesium and sodium sulfates which are both abundant in carbonaceous chondrites and highly soluble in water \citep{Kargel1991}. These brines are less dense than probable mixtures of ice and rock \citep{Kargel1991}, and therefore ought migrate to the icy satellite outer layers. Having a billion years to do so one expects subterranean oceans in icy satellites to be highly saline. \cite{Baland2014} used the observational data accumulated from the Cassini-Huygens mission to constrain the internal structure of Titan. They report that Titan very likely has a subterranean ocean with a bulk density of approximately $1.3$\,g\,cm$^{-3}$. Such a high density can be explained with a high concentration of solutes, especially salts. An aqueous solution of magnesium sulfate would require $1.88$\,molal of the latter to reproduce Titan's high oceanic bulk density \citep[see][and references therein]{Baland2014}. The expected ionic strength in Titan's subterranean ocean is thus much higher than for Earth's seawater. 

A super-Earth would differentiate much more rapidly than an icy satellite. For the case of a water planet, and under the high pressures prevailing at the water mantle to rocky core boundary \citep[see planetary internal parameter tables in][]{Levi2014}, a high pressure water ice layer ought quickly form. This layer would separate the ocean from the bulk of the rock, very probably having a significant effect on the water planet ocean's composition. Due to these reasons we will try to constrain the oceanic salinity of a water rich super-Earth from first principles, followed by an estimation for the ocean pH. 

In section $2$ we derive the thermal profile and dynamics of the water ice mantle, separating the ocean from the rocky interior. In section $3$ we estimate the possible concentrations of salt in the ocean, and describe the desalination pump (see subsection $3.2$). 
This is a fundamental and preliminary step to estimating the pH and carbonate speciation of the ocean. The speciation and the pH of the ocean depend on the chemical reactions in the ocean and on the charge balance criterion. What chemical reactions take place and in what way does CO$_2$ react with H$_2$O to maintain electrical neutrality depends on the concentration and identity of the salts dissolved in the ocean. The latter also determine the activity coefficients, required for the proper treatment of the chemistry taking place in the ocean. After we constrain the ocean's salinity we turn to formulate the equations governing the carbonate speciation in the ocean (in subsection $4.1$). In subsection $4.2$ we discuss the appropriate values to be used for the first and second dissociation constants of carbonic acid. In section $5$ we solve for the ocean's pH and carbonate speciation. Section $6$ is a discussion, and section $7$ is a summary.

\section{DYNAMICS OF THE ICE MANTLE}

Estimating the salinity of a water planet ocean requires the rate of transport of salt across the ice mantle. For this purpose we must first investigate the thermal profile and the dynamics of the ice mantle. In this section we will assume the ice mantle beneath the global ocean is composed of pure H$_2$O. Because the introduction of salt into the ocean may have a large effect on the melting curves of high-pressure ice polymorphs, then in addition to the pure water case we will also solve assuming the ocean has a salinity of $2.5$\,molal of NaCl. The latter is likely an upper bound value for a primordial unprocessed ocean, as we explain in the next section.   

The ocean floor is the upper thermal boundary layer of the ice mantle convection cell. If the thermal profile across this boundary layer, of width $\delta$, is conductive, one may write for it:
\begin{equation}
F_s=\kappa\frac{\Delta T}{\delta}
\end{equation}
Here $F_s$ is the heat flux at the ocean's floor, $\kappa$ is the thermal conductivity of either ice V or VI, and $\Delta T$ is the temperature difference across the boundary layer. From hydrostatic equilibrium the pressure difference across this boundary layer is:
\begin{equation}
\Delta P=\rho g\delta
\end{equation}
where $\rho$ is the density of the thermal boundary layer composition and $g$ the gravitational acceleration. From the last two equations we have the following temperature gradient:
\begin{equation}
\frac{\Delta T}{\Delta P}=\frac{F_s}{\kappa\rho g}\approx\frac{F_{s,Earth}X^{Si+Fe}}{\kappa\rho g_{Earth}}
\end{equation}
where on the right hand side we assume the heat flux is due to the radioactive decay in the rocky interior and scale it using values from Earth \citep[see eq.$23$ in][]{Levi2014}. The surface heat flux on Earth $F_{s,Earth}=0.087$\,W\,m$^{-2}$ \citep{Turcotte2002} and $g_{Earth}$ is the surface acceleration of gravity on Earth. $X^{Si+Fe}$ is the mass fraction of silicates and metals in our water-rich planet.

The temperature of the outer surface of this upper thermal boundary layer is set by the melting curve of the high pressure water ice polymorph, appropriate for the bottom ocean temperature. If the melting curve of this high-pressure ice has a gradient, $dT_m/dP$, then for
\begin{equation}
\frac{dT_m}{dP}<\frac{F_{s,Earth}X^{Si+Fe}}{\kappa\rho g_{Earth}}
\end{equation}
the conductive profile increases the temperature to values higher than the melting temperature. Due to the low viscosity of liquid water it would rapidly convect away the heat and cool. Thus, forcing the temperature profile in the boundary layer to follow the melting curve of the high pressure water ice. For the pure water system we adopt the melting curve formula for water ice VI from the IAPWS (revised release on the pressure along the melting and sublimation curves of ordinary water substance, September 2011) and use it to estimate an average value for the melting temperature gradient with pressure of $52$\,K\,GPa$^{-1}$. For the case where ice VI is in equilibrium with an ocean containing $2.5$\,molal of NaCl we adopt the melting curve from \cite{Journaux2013}, giving an average gradient of $36$\,K\,GPa$^{-1}$.  The density for water ice VI is taken from \cite{Bezacier2014} and its thermal conductivity is from \cite{Chen2011}. We find for the last condition, and for the pure water case, that:
\begin{equation}
X^{Si+Fe}>0.014
\end{equation}
For the salty ocean case a silicate and metal mass fraction of $0.01$ will be sufficient. 
Clearly, all of our studied planets have much higher silicate and metal mass fractions. Therefore, as the latter condition is a way to scale the heat flux, we assert, that the heat flux at the bottom of our studied oceans is sufficient to confine the upper boundary layer, underlying the ocean, to the melting curve of water ice VI. This may have important implications for the fractionation factor of salt between the ocean and the ocean floor solid matrix.

\begin{figure}[ht]
\centering
\includegraphics[trim=0.2cm 3.3cm 0.1cm 5cm , scale=0.6, clip]{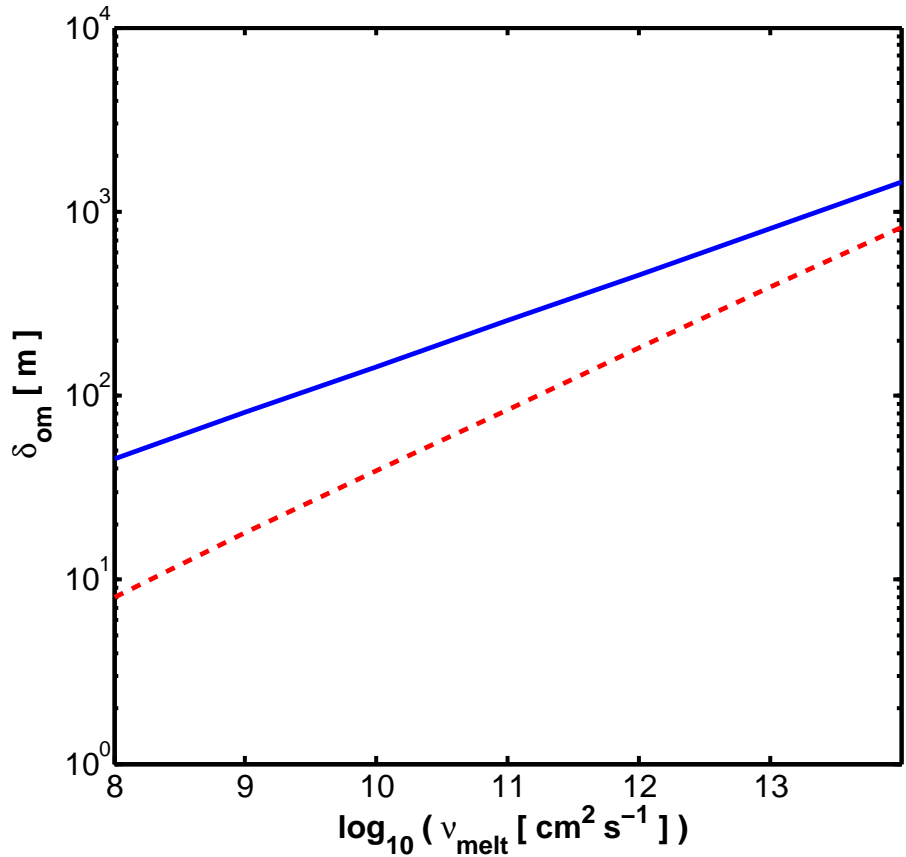}
\caption{\footnotesize{Thickness of the on-melt layer underlying the ocean versus a range for the viscosities of pre-melting ice VI. Solid blue curve is the solution for the pure water ocean. Dashed red curve is the solution for the case of the salty ocean.}}
\label{fig:OnMeltLayerWidth}
\end{figure}

As this upper thermal boundary layer is confined to the melting curve of ice VI, the viscosity contrast across it is probably low. This is because, phenomenologically, isoviscous contours on a temperature and pressure diagram have a topology similar to that of the melting curve \citep{poirier1985,durham1997}. Therefore, we assume the boundary reaches convective instability at a critical Rayleigh number of $1000$ \citep{Turcotte2002}. Experiments conducted on the rheology of ice VI do not constrain its strength very close to the melting temperature. Extrapolating on the temperature is highly unreliable when it comes to flow of crystalline solids. Hence, we solve for the width of this on-melt layer, $\delta_{om}$, for a wide range of viscosities (see results in fig.\ref{fig:OnMeltLayerWidth}). We note here that \cite{Poirier1981} reported a value of $1.8\times 10^{13}$\,cm$^2$\,s$^{-1}$ for the kinematic viscosity of ice VI at $10$\,K below melting conditions. 

The on-melt layer terminates at a depth where it reaches convective instability, and a adiabatic thermal profile with depth becomes more appropriate. However, a adiabatic thermal profile, in a pure water ice mantle, will tend to keep the temperature along the water ice mantle low \citep{Levi2014}. Consequently, the thermal profile will diverge very quickly from the melting curve of ice VI, resulting in a sharp increase in the strength of ice with respect to that within the on-melt layer. This increase in strength would favour heat conduction rather then convection. However, conduction would increase the temperature more steeply and towards melting conditions, causing the water ice to weaken. Therefore, underlying the on-melt layer is another layer of width, $\delta_{int}$, where the temperature is kept intermediate between the melting temperature of ice VI and the adiabatic temperature. We will refer to this layer as the intermediate layer. 

Here we approximate the temperature along the intermediate layer as the average between the adiabatic and the melting temperature. Adopting the rheological law for ice VI from \cite{durham1997}, 
we solve for the width of the intermediate layer by assuming it reaches convective instability. We adopt a higher critical Rayleigh number of $2000$, reflecting the somewhat higher viscosity contrast anticipated across this layer, in comparison to that across the on-melt layer.

Below the intermediate layer the thermal profile is approximated using an adiabat along the ice VII mantle. This adiabat terminates at a lower boundary layer separating the inner rocky core and the ice mantle. We refer the reader to \cite{Levi2014} for a in depth discussion and the development of the algorithm described here for estimating the thermal profile across the deep ice mantle of a water rich planet.    

The volume thermal expansivity and the density of water ice VI are taken from \cite{Bezacier2014}. The heat capacity adopted is from \cite{Tchijov2004}. The thermal conductivity is from \cite{Chen2011}. The viscosity of ice VI is from \cite{durham1997}.  In order to calculate the adiabat along the ice VII mantle the heat capacity and thermal expansivity for this high-pressure ice polymorph are needed, and are taken from \cite{fei1993}. The density of ice VII is taken from \cite{hemley87}.

Estimating the viscosity related to the non-Newtonian mechanisms reported in \cite{durham1997} requires an evaluation for the shear stress second invariant. We look for a consistent solution using iterations, first we adopt a value for this shear stress, then we solve for the thermal profile and mantle dynamics, we then scale the stress as \citep{Turcotte2002}:
\begin{equation}
\tau\approx\mu_m\frac{2u_0}{b}
\end{equation}
until a consistent solution is obtained. 
Here $\mu_m$ is the dynamic viscosity of the ice mantle, $u_0$ is the horizontal boundary velocity and $b$ is the depth of the convection cell. The resulting differential stress acting on the cross section of the upper thermal boundary layer is:
\begin{equation}
\sigma_b\approx\tau\frac{b}{\delta_{om}+\delta_{int}}
\end{equation}

\begin{figure}[ht]
\centering
\includegraphics[trim=0.2cm 4.3cm 0.2cm 5cm , scale=0.6, clip]{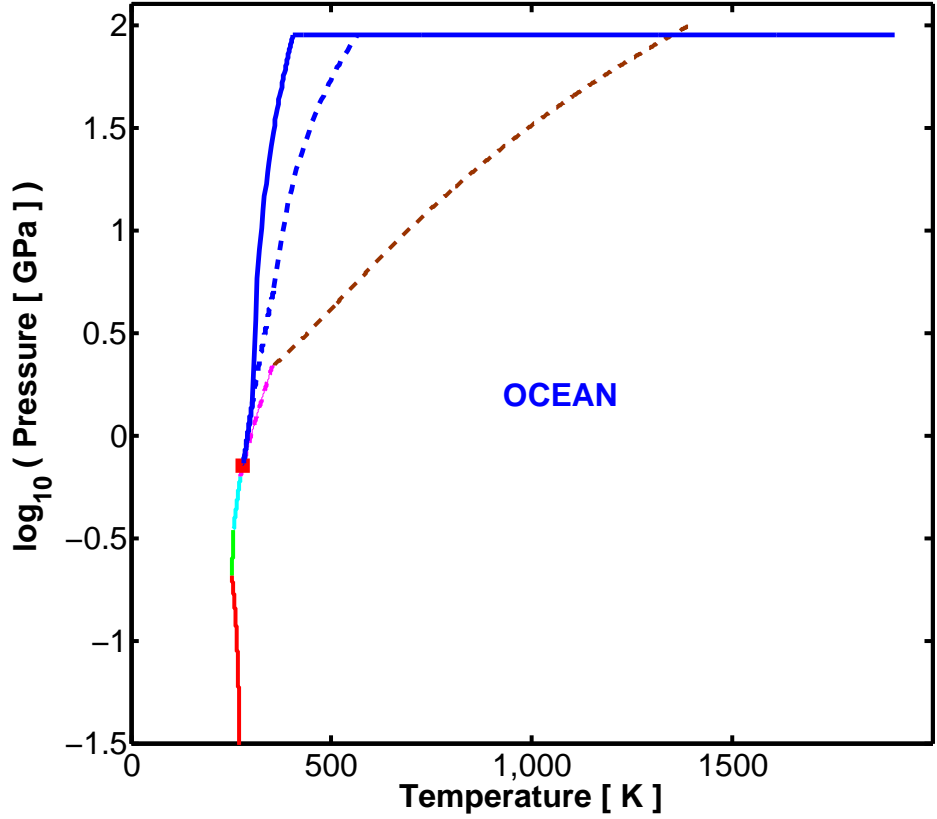}
\caption{\footnotesize{The thermal profile along the ice mantle, for the case of the pure ocean. Dashed blue curve is for the thick upper boundary layer solution. Solid blue curve is for the thin upper boundary layer solution. Red square marks a assumed condition at the ocean floor ($280$\,K and $0.7$\,GPa). Solid red curve is the melting curve of ice Ih. Solid green is the melting curve of ice III. Solid cyan is the melting curve of ice V. Dashed magenta is the melting curve of ice VI. These high pressure ice melting curves are from the IAPWS publication. Dashed brown is the melting curve of ice VII \citep{Lin2004,Lin2005}.}}
\label{fig:ThermalProfile}
\end{figure}

We adopt an aspect ratio of unity, which maximizes the Nusselt number. For these cells, assuming iso-viscosity, conservation of mass requires that the horizontal, $u_0$, and vertical, $v_0$, cell boundary velocities equal. We estimate these velocities as in \cite{Turcotte2002}, which give us the overturn time scale for the ice mantle:
\begin{equation}
t_{ot}=\frac{b}{v_0}
\end{equation} 
In table \ref{tab:2MEtable} we list the dynamic parameters we find for an ice mantle of a $2$M$_\Earth$ water planet with a $50$\% water mass fraction, for both the pure and salty ocean cases. In Fig.\ref{fig:ThermalProfile} we give the thermal profile across the water ice mantle, for the case of a pure water ocean. The thermal profile does not change substantially for the case of the saline ocean, for which the melting curve of ice VI is depressed. 

\begin{deluxetable}{ccccccccc}
\tablecolumns{8}
\tablewidth{0pc}
\tablecaption{Dynamical Parameters for a 2M$_\Earth$ Water Planet}
\tablehead{
\colhead{ocean} & \colhead{$\tau$}  & \colhead{$\delta_{int}$} & \colhead{Ra}  & \colhead{$v_0$}  & \colhead{t$_{ot}$} & \colhead{$\sigma_b$} &
\colhead{$\dot{M}_{liq}$}  \\
\colhead{composition} & \colhead{[MPa]} & \colhead{[km]} & \colhead{} & \colhead{[m\,s$^{-1}$]} & \colhead{[Myr]} & \colhead{[MPa]}  &
\colhead{[kg\,s$^{-1}$]} 
}
\startdata

\multirow{2}{*}{pure water}  & 6.1  &  60   &  3$\times 10^8$ & 3$\times 10^{-9}$ & 37  & 362 & 1.6$\times 10^{8}$   \\
                             & 8.5  &  378  &  9$\times 10^7$ & 1$\times 10^{-9}$ & 97  & 82  & 3.7$\times 10^{8}$   \\
\hline
\multirow{2}{*}{saline 2.5\,[molal]}      & 6.2  &  65   &  3$\times 10^8$ & 3$\times 10^{-9}$  & 39  & 344 & 1.6$\times 10^{8}$   \\
                                          & 8.0  &  366   &  1$\times 10^8$ & 1$\times 10^{-9}$ & 82  & 80  & 4.3$\times 10^{8}$    \\
\enddata
\tablecomments{Here it is assumed that $50$\% of the mass of the planet is H$_2$O, composing the water ice mantle and the ocean. $\tau$-shear stress in the ice mantle, $\delta_{int}$-thickness of the intermediate boundary layer, Ra-average Rayleigh number for the ice mantle, $v_0$-maximal vertical and horizontal velocity in the convection cell in the ice mantle, t$_{ot}$-mantle overturn time, $\sigma_b$-stress acting on the cross section of the upper thermal boundary layer, $\dot{M}_{liq}$-mass rate of melt entering the ocean. }
\label{tab:2MEtable}
\end{deluxetable}

It is a common practice in the field of plate tectonics research to compare between the differential stress, $\sigma_b$, and the ultimate tensile strength of the lithosphere, in order to assess the possibility of fracturing the lithosphere into plates.
Here, the upper boundary layer is under considerable hydrostatic pressure ($\sim 1$\,GPa) from the overlying ocean. The differential stress ought be considerably higher than the mean stress in order to activate fracture formation.  
Therefore, our derived values for $\sigma_b$ suggest the upper boundary layer experiences ductile deformation.

\begin{figure}[ht]
\centering
\includegraphics[trim=0.2cm 15.0cm 0.2cm 1cm , scale=0.7, clip]{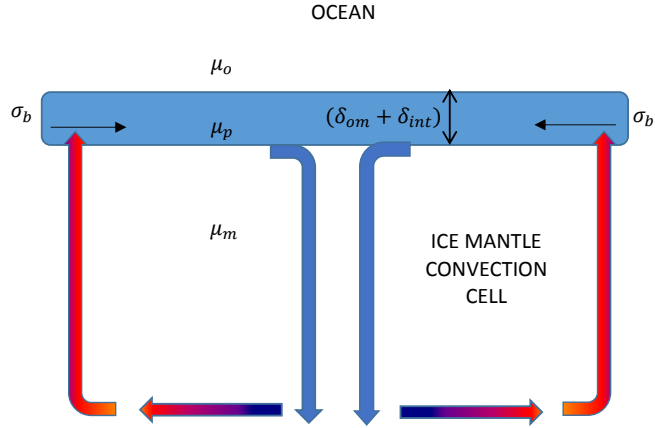}
\caption{\footnotesize{The upper boundary layer (shaded blue beam) of the ice mantle convection cell is under a large hydrostatic pressure due to the weight of the overlying ocean. The underlying convection cells apply a basal stress compressing the upper boundary layer. The state illustrated here is unstable against folding of the upper boundary.}}
\label{fig:FoldingInstability1}
\end{figure}

We approximate the upper boundary layer to be a viscous layer embedded in a viscous medium. This layer experiences compression due to traction from the underlying convection cells (see illustration in fig.\ref{fig:FoldingInstability1}). Such a scenario is unstable with respect to folding. \cite{Biot1957,Biot1959} investigated the evolution of sinusoidal folds in such a layer. He found that the deflection, $w$, of a fold of wavelength $\lambda$ increases with time as:
$$
w(x,t) = A_0e^{\frac{t}{\tau_f}}\cos\left(\frac{2\pi}{\lambda}x\right) 
$$ 
\begin{equation}
\frac{1}{\tau_f}\equiv\frac{\sigma_b}{\frac{1}{3}\mu_p(\delta_{om}+\delta_{int})^2\left(\frac{2\pi}{\lambda}\right)^2+\frac{\lambda}{\pi(\delta_{om}+\delta_{int})}(\mu_o+\mu_m)}
\end{equation}
where $\mu_p\sim 10^{19}$\,P, $\mu_m\sim 10^{23}$\,P and $\mu_o$ are the dynamic viscosities of: the folding layer, the underlying mantle and of the overlying ocean, respectively.     
The dynamic viscosity of the liquid ocean may be neglected here.

\begin{figure}[ht]
\centering
\includegraphics[trim=0.2cm 15.0cm 0.2cm 1cm , scale=0.7, clip]{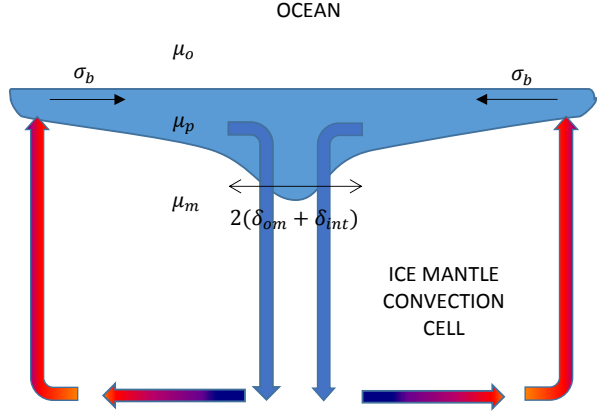}
\caption{\footnotesize{The upper boundary layer (shaded blue area) of the ice mantle convection cell is compressed due to traction from the underlying convection cells. Folds form in the upper boundary layer. Above regions of downwelling traction pulls the upper boundary layer inward. Therefore, generating ocean floor renewal.}}
\label{fig:FoldingInstability2}
\end{figure}

The deflection in the upper boundary layer above regions of downwelling is of particular interest. At these locations the deflected boundary will feel steady traction pulling it into the deep water ice mantle. Therefore, generating an effect analogous to slab pull, resulting in ocean floor renewal.
The width of this deflection is about $2(\delta_{om}+\delta_{int})$ as illustrated in Fig.\ref{fig:FoldingInstability2}. Thus, the wavelength of interest is:
\begin{equation}
\lambda\approx 4(\delta_{om}+\delta_{int})
\end{equation}
Knowing the wavelength we can estimate how the amplitude of the deflection increases with time.
For our system parameters, and assuming an initial deflection of $A_0=1$\,cm, the extent of the fold into the downwelling region reaches an amplitude of hundreds of kilometres after tens of Myr. This is the time scale needed to first initiate a continuous mechanism of ocean floor renewal. As this time scale is much shorter than the age of our studied planets we may infer that ocean floor renewal is active.      

In order to understand what happens at the upper boundary layer above upwellings, we note that the ability of the upper boundary layer to cool conductively is limited. This is due to the restrictions on the temperature gradient across the upper boundary layer as we have discussed above. In fact, heat lost to the ocean by conduction through the boundary layer can account for only a few percent of the radioactive budget. Therefore, the extra energy that is delivered by the hot ascending ice goes first into melting the upper boundary at that location.
Hence, above upwellings, the adiabatic profile of the convection cell may extend to the ocean floor. As is shown in Fig.\ref{fig:ThermalProfile}, this means that upwelled ice from the deep mantle would necessarily cross its melting curve. Therefore, feeding the ocean with hot positively buoyant liquid water. 

The global rate of melting may be approximated as:
\begin{equation}
\dot{M}_{liq}=\rho_{VI}WL_{GR}v_0
\end{equation}       
where $\rho_{VI}$ is the density of ice VI, $W$ is the width of the ridge at the point of melting and $L_{GR}$ is the global ridge length. From mass conservation $W$ is approximately equal to the width of the upper boundary layer. 

Assuming that the upper thermal boundary layer deforms into square plates of side $L$, and that each square contributes a ridge length of $2L$ to the total ridge length we have:
\begin{equation}
L_{GR}\sim\frac{4\pi R^2_p}{L^2}2L\approx\frac{8\pi R^2_p}{b}
\end{equation}
where $R_p$ is the planetary radius. On the right hand side of the last equation we assume a unit aspect ratio for the convection cells in the ice mantle.  

We find that there are two consistent solutions for the dynamics of the ice mantle, both for the pure water ocean and the saline ocean. These are tabulated in table \ref{tab:2MEtable}. The two thermal profiles along the ice mantle associated with the two solutions for the pure water scenario are given in Fig.\ref{fig:ThermalProfile}. One solution represents a thin and fast moving boundary layer, while the other solution is of a thick and sluggish upper boundary layer. We note that all solutions have similar global melting rates along the ocean floor ridges.

An additional point of interest is that for the model we have adopted, it seems that at the bottom boundary layer the thermal conditions cross the melting curve for pure water ice VII. This suggests the existence of a thin fluid layer right above the rocky interior at a pressure of about $90$\,GPa. A fluid layer between an internal rocky core and a high pressure ice mantle was suggested also by \cite{Noack2016}, though for planets with a very low water mass fraction. For the high water mass fractions we adopt here, this however may be an artifact of the assumption of a pure water mantle. The higher melting temperatures of filled ices may overrule such a phenomenon in our case of study \citep{Levi2013,Levi2014}.  

In this section we have solved for a $2$M$_\Earth$ planet assuming a water mass fraction of $50$\%. A lower water mass fraction, for a given planetary mass, means a higher heat flux due to the increased mass fraction of radioactive elements. Therefore, for lower water mass fractions convection in the ice mantle is likely more vigorous, and global melting rates at ridges are likely higher than those derived here.     

For a steady state the melting of ice at ridges along the ocean floor must be balanced by a solidification at equal rate of cold abyssal ocean water. Therefore, a \textit{cycle} of melting and solidification at the ocean floor is active. In the next section we examine its role in the  fractionation of salt between the global ocean and the deep ice mantle.

Finally, a steady state condition also implies that the melting of water ice along ridges, above upwellings, is not a major source of cooling, as the heat invested in melting is returned during solidification. Therefore, in steady state, cooling proceeds by the discharge of hot liquid water along ridges and into the ocean, and solidification of cold abyssal ocean water. For the heat capacity of liquid water from \cite{waite2007} and our calculated melting rates we find that the thin boundary solution can provide cooling at the rate of $10^{13}$\,W while the thick boundary solution can support cooling at the rate of $10^{14}$\,W. The radioactive heat flux integrated over the planetary surface is approximately $4\times 10^{13}$\,W. Since the latter is intermediate to the two former values we suggest the planet oscillates between the two tectonic regimes.

\section{OCEANIC SALT CONCENTRATIONS}

\subsection{Gravitational Limits on the Concentration of Salt}

Brine is much less dense than rock. Therefore, Earth's saline ocean stably floats on the oceanic lithosphere. For the case of a water-rich planet this may not be so, due to the much lower density difference between that of brine and of the high pressure water ice polymorphs that compose the ocean's bottom surface. Therefore, gravity may impose some restrictions on the possible maximum salinity of an ocean world.

\cite{Journaux2013} examined the melting point depression of ices VI and VII for four NaCl aqueous solution compositions of: $0.01$\,molal, $1$\,molal, $2.5$\,molal and $4$\,molal. They have noticed that when the brine concentration was increased from $1$\,molal to $2.5$\,molal there was an inversion in behaviour, where ice VI became less dense than the brine and floated in the chamber. However, Ice VII was denser than the brine for all examined samples. 

\begin{figure}[ht]
\centering
\includegraphics[trim=0.2cm 4.3cm 0.2cm 5cm , scale=0.55, clip]{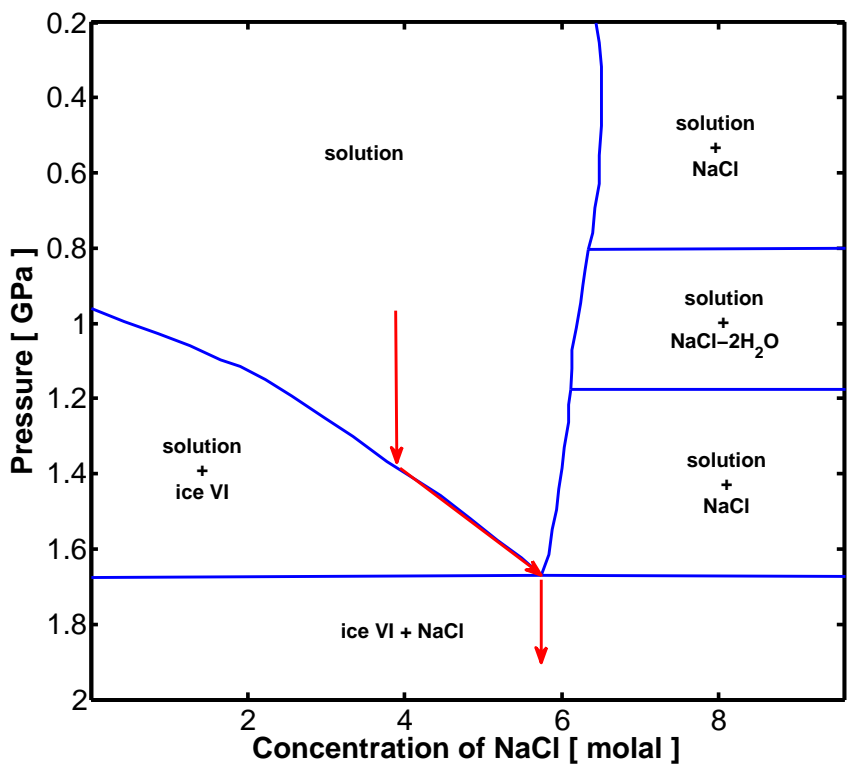}
\caption{\footnotesize{The phase diagram of the system H$_2$O-NaCl at room temperature \citep[reproduced from][]{Adams1931}}. See text for explanation of the red arrows.}
\label{fig:H2ONaClEutectic}
\end{figure}
            
In Fig.\ref{fig:H2ONaClEutectic} we reproduce the phase diagram for the binary system H$_2$O-NaCl at room temperature from \cite{Adams1931}.            
If during planetary formation strong water-rock interactions at high temperature resulted in a primordial ocean saltier than $1.75\pm 0.75$\,molal, then ice VII, rather then ice VI, would form the gravitationally stable ocean floor.
Fluctuations of density in the ocean would cause downward migration of some liquid parcels. In Fig.\ref{fig:H2ONaClEutectic} we give an example for the compositional evolution of such a liquid parcel (see red arrows). The sinking parcel will not change its composition until it reaches the liquidus. At which point grains of ice VI begin to form within the parcel. These grains have some NaCl dissolved within them, according to the solvus conditions, not shown in the figure. However, most of the salt remains dissolved in the liquid parcel. Therefore, the now denser parcel continues to sink, its composition restrained to the liquidus. On the other hand, the ice VI grains are positively buoyant and thus migrate upward, crossing their melting condition producing a less saline upper ocean layer. The liquid parcel sinks while following the liquidus until reaching the eutectic composition. Beyond that point its salt content largely forms grains of halite, whose high density (see Fig.\ref{fig:SaltVIceDensity}) ensures they descend down to the ice VII layer. Ice mantle overturn then sediments these pure salt grains on the ice-rock mantle interface.        

This process is likely vary efficient in removing salt from the ocean, if indeed its initial salinity was high. When the ocean's average salinity drops below $1.75\pm 0.75$\,molal the ice VI grains become negatively buoyant in the background ocean. Therefore, an ice VI layer begins to form above the ice VII layer. Consequently, the ocean desalination process as described here ceases. 
This gravitational upper limit on the oceanic salt concentration is however not very stringent, in the sense that it is higher than the concentration of NaCl in Earth's seawater which is approximately $0.5$\,molal \citep[in Earth's seawater Na$^+$ and Cl$^-$ concentrations are $0.47$\,molal and $0.55$\,molal, respectively, see][]{zeebe2001}. 
It is interesting to note that the salt concentration suggested for Titan \citep[$1.88$\,molal,][]{Baland2014} falls within this range of allowed maximum concentration.  

The criterion of gravitational stability of the ocean may prove to be even more restrictive. One ought consider the proliferation of clathrate forming chemical species, both dissolved in the ocean, and embedded within its bottom surface ice layer (\cite{Levi2014,Levi2017}). Thus, the ocean's gravitational stability requires consideration of the densities of clathrates as well. We have compared between the densities of aqueous NaCl solutions \citep{Lvov1990} and of the CO$_2$ structure I clathrate hydrate \citep{Levi2017}. We find that at $273.15$\,K and $0.8$\,GPa their densities become equal for a $3.14$\,wt\% ($0.56$\,molal) of NaCl. This is about the value of Earth's seawater. Therefore, the richer the ocean floor is in CO$_2$ clathrate hydrates the less likely it is that a water planet ocean was ever saltier than Earth's oceans.

\subsection{The Desalination Pump}

The constraints placed by gravity on the upper bound values for the ocean's salinity are not very limiting. In addition, whether these upper bound values can be reached depends on the outward transport of solutes along the evolutionary path of the ocean exoplanet. Such an exploration is beyond the scope of this work. However, it may be that the melting and solidification cycle discussed in the previous section can further constrain the ocean's salinity. This requires an examination of the high pressure experimental data available for salt-water solutions. 

\cite{Frank2006} studied the H$_2$O-NaCl system up to $30$\,GPa at room temperature using a diamond anvil cell. A pressure of $30$\,GPa would reflect the conditions prevailing at the bottom of an ice mantle of a $2$M$_\Earth$ planet with an approximately $15$\% ice mass fraction \citep[see water planet internal structure tables in][]{Levi2014}. 
These authors compressed two NaCl aqueous solutions: one of $1.6$\,mol\% ($0.9$\,molal) and the other of $3.2$\,mol\% ($1.9$\,molal) of NaCl. In the more dilute case the diffraction lines could be indexed using ice VII alone. In the more solute rich case there were clear indications for the formation of halite. The authors concluded that, at room temperature, the many natural voids available within the ice VII structure could host Na$^+$ and Cl$^-$ ions with a solubility of $2.4\pm 0.8$\,mol\% ($1.4\pm 0.5$\,molal). 
\cite{Daniel2008} give a brief report over their investigation of the H$_2$O-NaCl system, and their results do not contradict the high solubility found by \cite{Frank2006}, and the analysis of this work.

\cite{Adams1931} measured the volume of the liquid binary solution H$_2$O-NaCl for various compositions and room temperature. In his experiment he investigated pressures up to the transition to ice VI. From the partial volumes he derived chemical potentials, and inferred the solubility of NaCl in liquid water in the pressure range of $1$\,bar to $1.6$\,GPa. At $0.3$\,GPa he estimated the solubility to be $6.52$\,molal.  
\cite{Sawamura2007b} yielded a solubility in liquid water of $6.61$\,molal of NaCl, at these same pressure and temperature conditions.
\cite{Adams1931} further reported that the solubility varies very little with pressure, and derived a value for the solubility of $6.3$\,molal at $1$\,GPa, i.e. at the bottom of a water planet ocean.   

When ionic solutes become embedded in the water ice lattice they ought influence their surrounding water molecules. Based on their X-ray diffraction data and Raman spectroscopy of the O-H stretching frequency, \cite{Frank2006} suggested the water hydrogen atoms tend to align themselves closer towards the Cl$^-$ ion, and also experience partial ordering. The increased order, around the ionic solute, is manifested by the protons becoming equidistant from the oxygen atoms, thus resembling the structure of ice X. Therefore, for temperatures above room temperature, and in the stability regime of water ice VII, the protons may resist this ordering with implications for the solubility of salt in high pressure water ice polymorphs. In order to test their findings \cite{Frank2008} repeated their earlier experiments, though now for temperatures higher than room temperature. The authors prepared a $1.6$\,mol\% ($0.9$\,molal) aqueous solution of NaCl at room temperature which again was compressed to different pressures up to about $30$\,GPa. As before halite was not observed at room temperature. When the samples were heated to $500$\,K halite diffraction lines began to appear and intensified with the increasing temperature. The appearance of halite beyond $500$\,K was reported not to be very sensitive to the pressure. The halite signature disappeared once more after the samples crossed the melting point for the given isobar, meaning the formed fluid was not saturated. The authors concluded that at $500$\,K the salt began to exsolve out of the ice. We refer the interested reader to \cite{Bove2015} for more details on the effects of the inclusion of salt on the structure of water ice.

From Fig.\ref{fig:ThermalProfile} we see that for the case of the thin upper boundary layer, a temperature of $500$\,K is exceeded only within the lower boundary layer, separating the inner rock core from the ice mantle. For the case of the thick upper boundary layer this temperature is exceeded at a pressure level of about $53$\,GPa along the ice mantle adiabat. These are likely upper bound values for the salt exsolution pressure level. As we have shown in \cite{Levi2014}, an ice mantle enriched in filled ice would have a higher temperature than along a pure water ice mantle.

\begin{figure}[ht]
\centering
\includegraphics[trim=0.2cm 4.3cm 0.2cm 5cm , scale=0.55, clip]{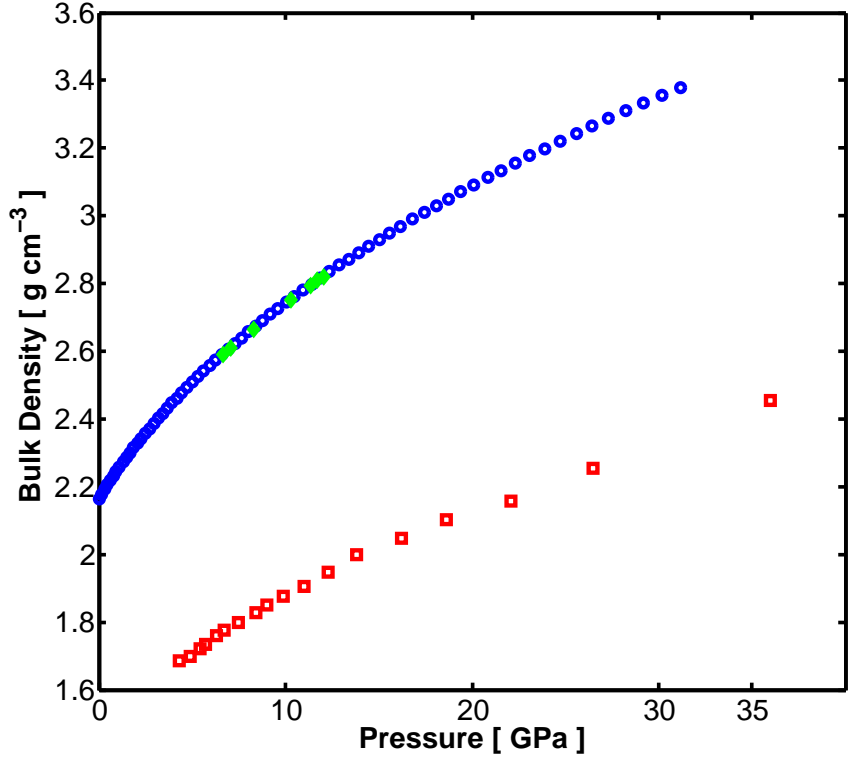}
\caption{\footnotesize{Experimental data for the variation of the density with pressure at $300$\,K, for NaCl and ice VII. Blue hollow circles are densities of NaCl from \cite{Decker1971}. Green diamonds are densities of NaCl from \cite{Matsui2012}. Red hollow squares are densities for pure ice VII from \cite{hemley87}.}}
\label{fig:SaltVIceDensity}
\end{figure}

These experimental results indicate that during the solidification of ocean water, in the melting-solidification cycle explained above, salts in the form of hydrated ions from the ocean can become incorporated into the continuously forming ocean floor. In the process of convective overturn of the ice mantle, thermal conditions in the downwelling arm of the convection cell are sufficient to promote the unmixing of the solid solution into pure ice VII and pure salt. In Fig.\ref{fig:SaltVIceDensity} we plot the density of pure NaCl versus that for pure water ice VII at room temperature and up to $30$\,GPa. It is clear that under these conditions the NaCl salt grains are much denser than ice VII. Therefore, the salt grains formed following this exsolution process sediment out of the ice mantle and onto the rocky interior.  

Water planets as considered here, have a water mass fraction of a few tens of percent, and pressures of tens of GPa at the bottom of their ice mantles.
Temperatures at the rocky core and ice mantle boundary far exceed the conditions for the exsolution of salt (see Fig.\ref{fig:ThermalProfile}).  Therefore, under such conditions it is probable that salts from the rocky core cannot incorporate into the overlying ice mantle as dissolved ions, to be transported outward. 

For the case of icy satellites, a mechanism that is often suggested for the outward transport of salt are pockets of brine \citep[e.g.][]{Choblet2017,Klara2018}. For such small icy bodies these pockets are positively buoyant, that is since their surrounding is a partially differentiated layer of ice and rock. Partial differentiation of ice from rock is not likely for a planet. Therefore, outward migration of a pocket of brine from the rock-ice mantle boundary requires it to be less dense than the surrounding ice. From \cite{Goncharov2009} we know that the density of water ice at the range $2-60$\,GPa is at most a factor of $1.15$ larger than that of the pure fluid. The equation of state for high pressure brine as a function of composition is not known. Here we approximate the contribution of dissolved NaCl to the solution density using low pressure data from \cite{Lvov1990}. They find that adding more than about $20$\,wt\% (i.e. an abundance of about $0.07$) of NaCl to liquid water can increase the solution density by more than a factor of $1.15$. Therefore, we tentatively assume, that for a brine pocket to be positively buoyant it should not contain a molar abundance of dissolved NaCl in excess of about $0.07$.

A pocket of brine forms if local conditions coincide with melting conditions. 
The thermal profiles in the ice mantles of water planets, as described in Fig.\ref{fig:ThermalProfile}, show that the difference between the adiabatic temperature and the ice melting temperature increases substantially with the increasing pressure. Therefore, in order to produce local melting conditions added solutes ought to lower the melting temperature by hundreds of degrees. To test whether this is possible we derive from \cite{Goncharov2009} the entropy of fusion for ice VII, for the pure water system, from $2$\,GPa to $60$\,GPa. The usage of the pure water system data is explained in appendix A.

For pressures higher than $45$\,GPa:
\begin{equation}
S_f = 3\times 10^{-6}P^3-0.00046P^2+0.024P-0.4 \quad [kJ\,K^{-1}\,mol^{-1}]
\end{equation}
and, for pressures lower than $45$\,GPa:
\begin{equation}
S_f = 2.83\times 10^{-5}P^2+0.00082P+0.026 \quad [kJ\,K^{-1}\,mol^{-1}]
\end{equation}
where $P$ is in GPa.
We use the following equation to estimate the melting point depression due to the addition of salt (see appendix A for the derivation):
\begin{equation}\label{MeltDepressionForm}
-S_f\Delta{T}+\frac{1}{2}\left(\frac{\partial S_f}{\partial T}\right)(\Delta{T})^2=k(T_0-\Delta{T})\ln\left(\gamma_{H_2O}(1-X_{NaCl})\right)
\end{equation}
where $\Delta{T}$ is the melting point depression, $k$ is Boltzmann's constant, $T_0$ is the undepressed melting temperature, $\gamma_{H_2O}$ is the activity coefficient of water and $X_{NaCl}$ is the abundance of salt dissolved into the fluid solution.

\begin{figure}[ht]
\centering
\includegraphics[trim=0.2cm 4.3cm 0.2cm 5cm , scale=0.55, clip]{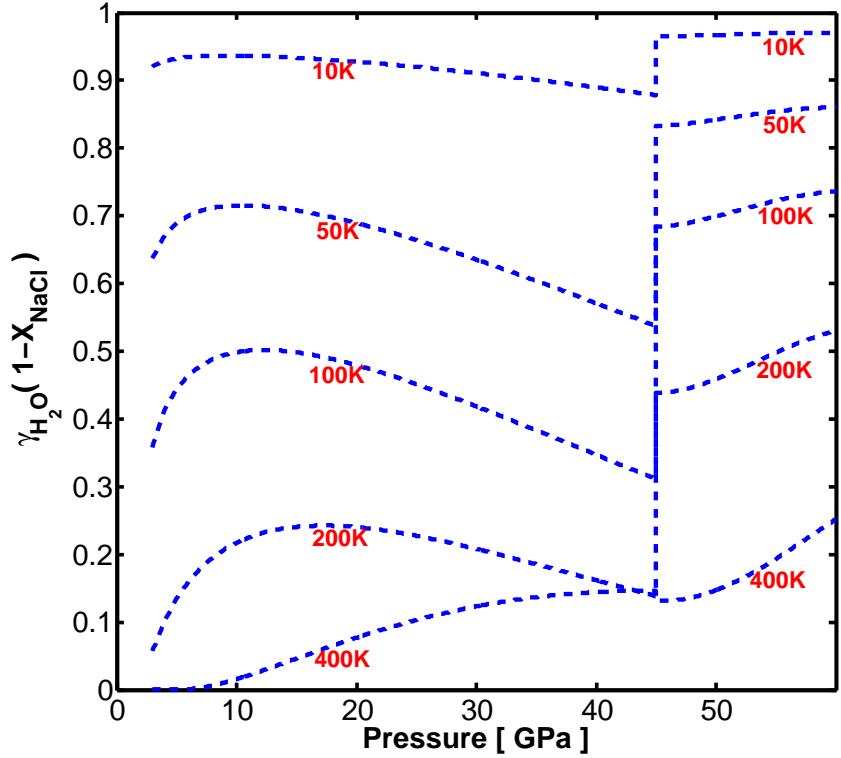}
\caption{\footnotesize{Iso-melt depression curves for ice VII.}}
\label{fig:MeltDepression}
\end{figure}

In Fig.\ref{fig:MeltDepression} we plot various iso-melting temperature depression curves.
From the figure we see that lowering the melting temperature by $400$\,K requires:
\begin{equation}
\gamma_{H_2O}(1-X_{NaCl})< 0.1
\end{equation} 
which yields for our adopted maximum salt concentration of $X_{NaCl}=0.07$ the following restriction on the activity coefficient:
\begin{equation}\label{ActivityCriterion}
\gamma_{H_2O}< 0.11
\end{equation}
\cite{Journaux2013} report on the melting point depression of ice VII when formed from a solution with a $19$wt\% (or abundance of $0.067$) of NaCl. We infer from their melting point depression data that the activity coefficient of water varies from $0.76$ to $0.85$, when increasing the pressure from $3$\,GPa to $6$\,GPa.
\cite{Frank2008} found for a $5$\,wt\% NaCl-water solution a melting point depression of $40$\,K at $25$\,GPa. We find this corresponds to an activity coefficient for water of $0.74$. We note that the latter activity coefficient will only allow for a maximum melting point depression of the order of tens of Kelvin, for $X_{NaCl}=0.07$. Therefore, both conditions: melting, and positive buoyancy of the brine pocket, are not met simultaneously.   

\begin{figure}[ht]
\centering
\includegraphics[trim=0.2cm 4.3cm 0.2cm 5cm , scale=0.55, clip]{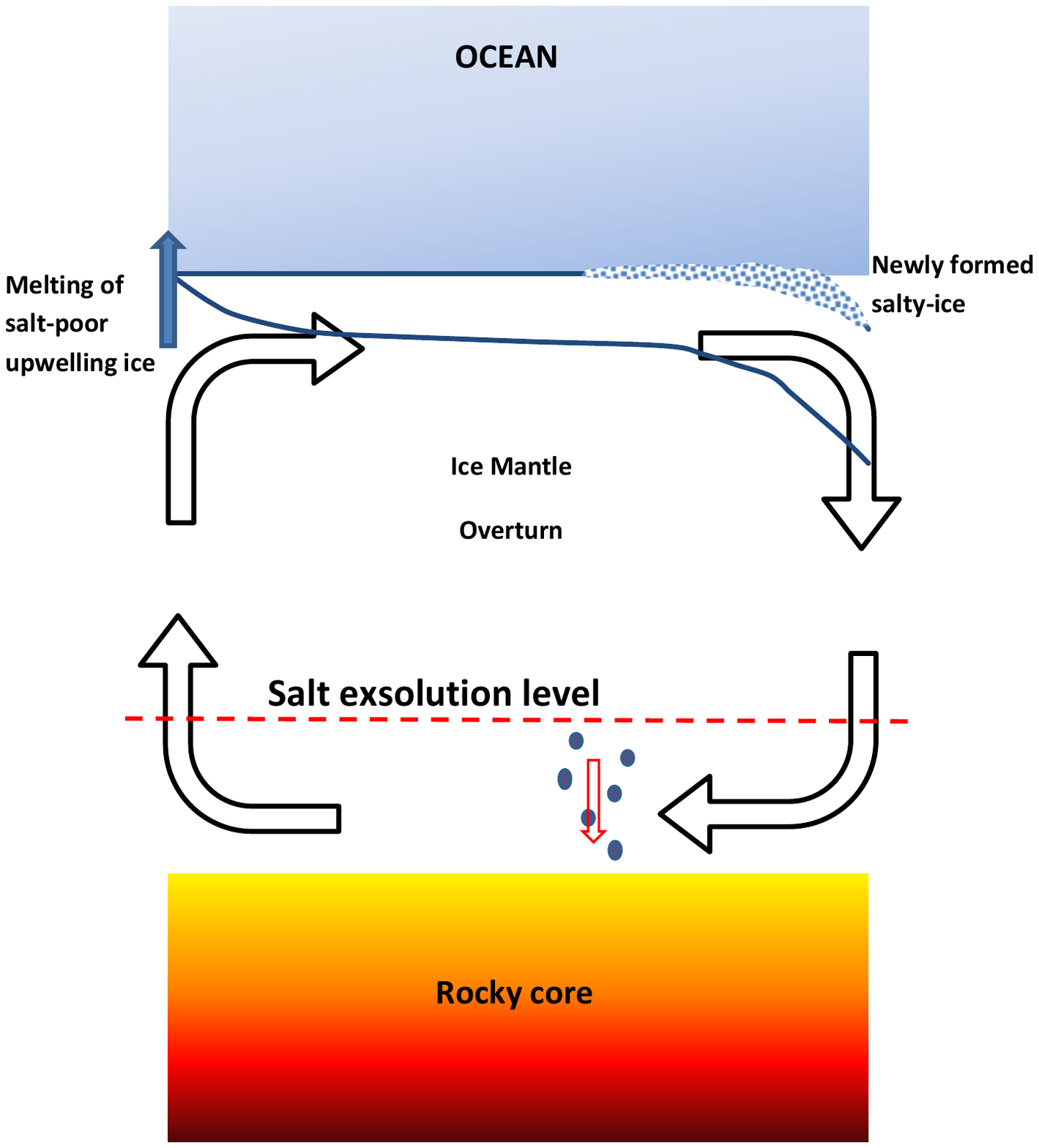}
\caption{\footnotesize{An illustration of the oceanic desalination pump active in water-rich planets. High-pressure ice, devoid of salt, upwells and melts below ridges. Consequently, an ocean in steady-state experiences solidification of liquid at the same rate. This solidification partitions hydrated ions of salt out of the liquid ocean, and into the solid ocean floor. Overturn of the ice mantle causes the newly formed ocean floor to downwell. In the deep ice mantle the interstitial ions are exsolved out of the water ice matrix. The formed grains of salt sediment onto the rocky core.}}
\label{fig:DesalinationPump}
\end{figure}

If salt is partitioned out of the ocean in the downwelling arm of the ice mantle overturn, while the upwelling arm is prevented from replenishing the ocean, a \textit{pump} is established. This pump would tend to desalinate the ocean over time. The pump is illustrated in Fig.\ref{fig:DesalinationPump}. 

We turn to estimate the rate of oceanic desalination, as enforced by this pump.
In steady-state the mass of the ocean, $M_{oc}$, is constant. Therefore, the rate of melting at spreading centres equals the rate with which ocean water solidify (see $\dot{M}_{liq}$ in section $2$).
We separate the solidification into a fraction $\chi$ forming a new ocean floor, and a fraction $1-\chi$ representing macroscopic entrapment of ocean water in regions of plate bending.   
Hydrated ions of salt dissolved in the ocean are incorporated into the newly forming ocean floor with some fractionation factor, $f_1$. Direct entrapment where the plates bend would fractionate ions into the ocean floor with a fractionation factor of $f_2$.   

The average salt concentration in the ocean is $m_{salt}$ (in molal unit). In \cite{Levi2017} we have shown that winds and tides are insufficient for efficiently mixing the deep oceans we study here. Therefore, the bottom of the ocean may be enriched in salt relative to the average ocean salinity. In appendix B we estimate this factor of enrichment to be at least $2.5$. 
If an oceanic mass element near the ocean floor, $dM_{oc}$, solidifies, the number of moles of salt partitioned out of the ocean is:
\begin{equation}
dn_{salt}\approx 2.5dM_{oc}\chi f_1m_{salt} + 2.5dM_{oc}(1-\chi) f_2m_{salt}
\end{equation}
If $n_{salt}$ is the total number of moles of salt dissolved in the ocean, we have:
\begin{equation}
n_{salt} = M_{oc}m_{salt}
\end{equation}
Differentiating with respect to time, and using the steady-state assumption for the ocean mass, yields:
\begin{equation}
\frac{d}{dt}m_{salt}\approx -2.5\frac{\dot{M}_{liq}}{M_{oc}}m_{salt}\left[\chi f_1+(1-\chi)f_2\right]
\end{equation}
Taking $t=0$ as the time when the ocean may have attained maximum salinity, as mandated by the criterion of gravitational stability, we can solve the last equation to obtain:
\begin{equation}\label{desalinationtime}
\ln\left(\frac{m_{salt}\left[mol\,kg^{-1}\right]}{0.56}\right)=-\frac{t}{\tau_{d}}
\end{equation} 
where the ocean desalination time scale is:
\begin{equation}
\tau_{d}\approx\frac{1}{2.5\left[\chi f_1+(1-\chi)f_2\right]}\frac{M_{oc}}{\dot{M}_{liq}}
\end{equation}

The region where the upper boundary layer (see fig.\ref{fig:FoldingInstability2}) folds experiences both an enhanced compression and tension. The strain in the fold can be written as:
\begin{equation}
\epsilon=\frac{y}{R_c}
\end{equation}
where $y$ is the normal distance from the neutral axis, and $R_c$ is the radius of curvature of the fold. Assuming that the neutral axis lies along the center of the boundary layer, the maximal value for $y$ is:
\begin{equation}
y_{max}=\frac{\delta_{om}+\delta_{int}}{2}
\end{equation}  
Thus, the differential stress in the fold is \citep{Turcotte2002}:
\begin{equation}
\sigma_{fold}=\frac{Y}{1-\nu^2}\frac{y}{R_c}
\end{equation}
Here $Y$ is the Young's modulus of water ice VI. Single crystal \citep{Shimizu1996,Tulk1996} and polycrystalline \citep{Shaw1986} experiments assign the latter a value of about $18$\,GPa. For the Poisson's ratio, $\nu$, we adopt a value of $0.32$ \citep{Tulk1996}. 

The functional form for the radius of curvature, $R_c$, for various flow regimes, is still debated. For the purpose of studying Earth's internal dynamics, \cite{Crowley2012} adopted a value of $R_c/b=0.1$, where $b$ is the depth of the convection cell. \cite{Rose2011} and \cite{Korenaga2010} reported that the value of $R_c/b$ decreases when the Rayleigh number considered increases. The total mantle Rayleigh number for our case of study is larger than the value for Earth's mantle \citep{Crowley2012}. Therefore, by adopting a value of $0.1$ we may be overestimating the length scale of the radius of curvature. This results in a lower estimate for the stresses in the fold.

For the values given above we find that the maximal stress in the fold is $10.3$\,GPa and $1.6$\,GPa for the thick and thin upper boundary solutions respectively. 
On Earth, the intense differential stress at plate bending, during subduction, promotes both brittle failure and plastic creep \citep{Billen2005}. This is often manifested as earthquakes, capable of cutting through a substantial part of Earth's $\approx 100$\,km lithosphere \citep{Kikuchi1995}. If similar activity occurs for our case of study, then ocean water my be forced into the downwelling oceanic floor. Consequently, the injected ocean water will follow the liquidus (see fig.\ref{fig:H2ONaClEutectic}), while increasing its density. It is probable that such a direct and macroscopic injection of ocean water would result in a fractionation factor of $f_2=1$, regardless of the identity of the hydrated ionic species. 
           
However, away from regions of plate bending, the fractionation of ions from the ocean into the continuously forming ocean floor is microscopic in its nature. 
Therefore, the fractionation factor $f_1$ ought depend on the rate of ocean water solidification. If the solidification front advances fast enough, kinetics determines its value.
As we have discussed above, the criterion for the gravitational stability of the ocean gives a maximum salinity of about $0.56$\,molal. This is less than the solubility of NaCl in high pressure ice. Therefore, if solidification is fast, the ions dissolved in the liquid, are easily incorporated as interstitials within the high-pressure ice, without supersaturating it, and $f_1=1$.   
If, on the other hand, the solidification front advances slowly, thermodynamics controls the value of the fractionation factor. In this case, equilibrium is attained, after a fraction $1-f_1$ of the dissolved ions diffuse out of the solidifying mass, and remain in the ocean.         
For this case, we estimate the fractionation factor by taking the ratio between the solubility of Na$^+$ and Cl$^-$ in the phase composing the ocean floor and the overlying pressurized liquid water.

The solidification front advances in the ocean with a velocity:
\begin{equation}
v_{sf}\sim\frac{\dot{M}_{liq}}{\rho_{oc}4\pi R^2_p}\sim 1\,[\AA\, s^{-1}]
\end{equation}
where $\rho_{oc}$ is the density of the ocean. This is a lower bound value for the solidification front velocity because we assume the solidification takes place on the entire ocean floor.  
The diffusion time-scale of ions in water is:
\begin{equation}
t_{dif}\sim\frac{l^2}{D_{ion}}
\end{equation} 
The diffusion coefficient of NaCl ions in water is $10^{-5}$\,cm$^2$\,s$^{-1}$ \citep{Li1974}. Advancing a distance $l=100\AA$ takes $10^{-7}$\,s, whereas it takes $100$\,s for the solidification front to cover the same distance. Therefore, the kinetics of solidification is slow enough to let $f_1$ attain its equilibrium value.     

From the experimental data given above, we obtain for the fractionation factor a value of $f_1\approx 0.22\pm 0.08$. This value is derived by adopting the experimental solubility for NaCl in ice VII. The solubility is lower for the case of ice VI. The latter is the more appropriate water ice structure for the case we are solving for, in which the ocean floor temperature is low. \cite{Journaux2017} report on the solubility of RbI in ice VI. They further report on the melting point depression in the H$_2$O-RbI binary system. They find that the melting point depression is similar to that in the H$_2$O-NaCl system. We derived the activity coefficients for NaCl and RbI in an aqueous solution, which can be translated to the osmotic coefficient for the solvent (water). The latter is easily related to the activity of water in the binary solution \citep{Blandamer2005}. We find that the activity of water differs by less than $1$\% between the two salts. Therefore, the similar melting point depression can also be translated to a similar solubility.   
At about $10$\,K below the melting temperature \cite{Journaux2017} find a fractionation factor $f_1\approx 0.005$. 
For $2-3$\,K below the melting temperature they give an upper bound value for the fractionation factor of about $0.3$. Since the latter is an upper bound value rather then a measurement for the solubility we shall adopt here their reported value of $0.005\pm 0.002$.  We refer the interested reader to \cite{Journaux2017} for a discussion on the difficulty in measuring the fractionation factor close to melting conditions, as in the on-melt layer (see discussion on the on-melt layer in section $2$).        

In \cite{Levi2014} we have derived planetary radii for a $2$M$_\Earth$ planet, composed of various ice mass fractions. For a $50$\% ice mass fraction we calculated a planetary radius of $R_p=9480$\,km. 
For an ocean depth of $80$\,km, the ocean mass is $M_{oc}\approx 9\times 10^{22}$\,kg.
For the melting rates tabulated in the previous section (adopting the average value between the thin and thick boundary layer solutions), the desalination time scale is:
\begin{equation}
\tau_{d}\sim\frac{4.3}{\chi (f_1-1)+1} \quad [Myr]
\end{equation}

If there is no macroscopic entrapment of oceanic water at regions of plate bending then $\chi=1$. For a planet with an ocean floor temperature that can stabilize ice VII we find $\tau_d\approx 20$\,Myr. For an ocean floor composed of ice VI we have $\tau_d\approx 860$\,Myr. If macroscopic entrapment can account for $1$\% of the reprocessing of ocean water (i.e. $\chi=0.99$) than the latter reduces to $\tau_d\approx 290$\,Myr. 
The median age of transiting planets' host stars is $\sim 5$\,Gyr \citep{Bonfanti2016}. Therefore, the desalination mechanism is efficient in removing salt out of the ocean, with fundamental implications for the habitability of our studied worlds.  

\begin{figure}[ht]
\centering
\includegraphics[trim=0.2cm 4.3cm 0.2cm 5cm , scale=0.55, clip]{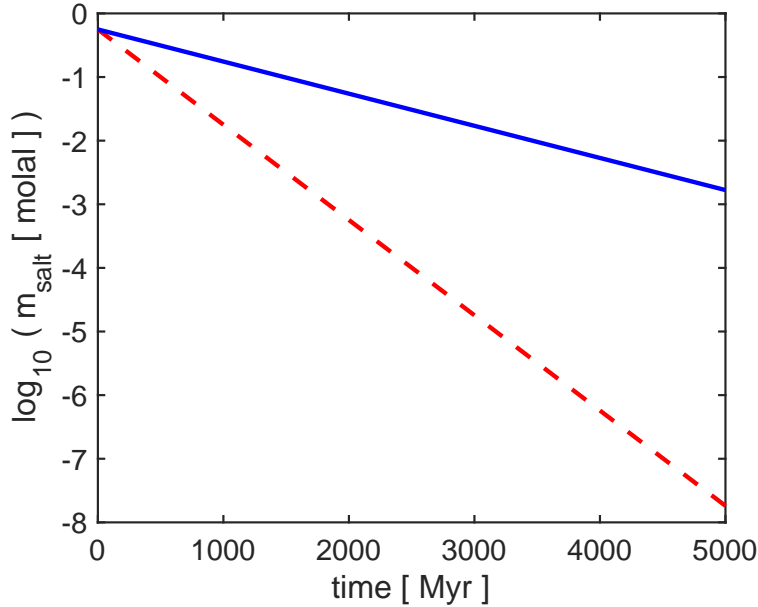}
\caption{\footnotesize{Evolution of the concentration of salt with time. At $t=0$ the maximum salinity is about $0.56$\,molal, due to gravitational constraints. Solid (blue) curve is for a desalination time scale of $860$\,Myr, and dashed (red) curve is for a time scale of $290$\,Myr. }}
\label{fig:SaltConcTime}
\end{figure} 

In Fig.\ref{fig:SaltConcTime} we plot the oceanic salinity versus time. After $5$\,Gyr the maximal concentration is of the order of millimolal.
Under these circumstances we conclude that ocean worlds, which are likely rich in volatiles, have oceans which are probably poor in salts.

\section{GOVERNING CHEMICAL EQUATIONS}

A fundamental difference exists between: (1) an ocean with a rocky bottom (i.e. Earth's oceans), and (2) the ocean of a water-rich exoplanet (where the ocean floor is composed of high pressure ice polymorphs). The first is in general a \textit{closed} system with regards to the abundance of carbon, whereas the second represents an \textit{open} system. 

An open system can exchange CO$_2$ with its environment. For example, a cup of water, or the surface of Earth's ocean, can exchange CO$_2$ with Earth's atmosphere. For these two examples the Earth's atmosphere is an infinite reservoir of CO$_2$, that can keep them saturated. In other words, if some fraction out of the freely dissolved CO$_2$ reacts to form carbonate or bicarbonate, there exists enough CO$_2$ in the reservoir (i.e. the atmosphere) to replenish what was lost, and bring both systems back to saturation. Therefore, the solubility of freely dissolved CO$_2$ is independent of the system's pH. For the two examples mentioned, the solubility of freely dissolved CO$_2$ depends only on the partial pressure of atmospheric CO$_2$, according to Henry's law.
 
However, Earth's atmospheric CO$_2$ is not an infinite reservoir for the bulk of Earth's ocean. Since most of Earth's carbon was deposited into rock, the total amount of CO$_2$ available for the ocean is limited. Therefore, the total freely dissolved CO$_2$ in the ocean is what is left of the total dissolved inorganic carbon, after subtracting the formed carbonate and bicarbonate. This means that in a closed system the abundance of freely dissolved CO$_2$ becomes pH dependent.
 
Water-rich exoplanets are probably also very rich in CO$_2$. Solid state convection within the ice mantles of these planets results in the outgassing of CO$_2$ into their oceans. These oceans saturate, and attain CO$_2$ concentration levels appropriate for an equilibrium with SI CO$_2$ clathrates, the latter composing the ocean floor. Freely dissolved CO$_2$ that reacts to form carbonate and bicarbonate may be replenished by the dissociation of SI CO$_2$ clathrates from the ocean floor. If the freely dissolved CO$_2$ in the ocean exceeds the solubility in equilibrium with its clathrate phase, grains of clathrate form and restore the equilibrium levels of CO$_2$. We refer the interested reader to \cite{Levi2017} for an analysis of the CO$_2$-H$_2$O system in ocean worlds. The behaviour described here is indicative of an open system, one in which the abundance of freely dissolved CO$_2$ is independent of the oceanic pH level. Rather, it is the oceanic abundance of the freely dissolved CO$_2$ that controls the ocean's pH and the carbonate system speciation.

\subsection{The Speciation Equations}

The equilibria relations between the carbonate species are \citep{zeebe2001}:
\begin{eqnarray}
CO_2+H_2O \rightleftarrows HCO_3^{-} + H^{+} \\
HCO_3^{-} \rightleftarrows CO_3^{2-} + H^{+}
\end{eqnarray}
Here CO$_2$ represents freely dissolved carbon-dioxide plus a trace amount of true carbonic acid, H$_2$CO$_3$. For our pressure and temperature range  of interest $H^{+}$ represents a hydrated proton, which is part of a larger complex \citep[an unbonded proton below $20$\,GPa has a very short lifetime,][]{Goncharov2009}. Two such commonly discussed complexes are: the Zundel (H$_5$O$_2^{+}$) structure, where the proton is shared by a pair of neighbouring water molecules and the Eigen (H$_9$O$_4^{+}$) structure for which a hydronium (H$_3$O$^{+}$) core is bonded to three water molecules \citep{Markovitch2007}. \cite{Marx1999} have shown that many complexes exist between which the hydrated proton transfers. Therefore, the Zundel and Eigen structures should only be considered as limiting cases.  

In equilibrium the relations between the concentrations of the different carbonate species are:
\begin{equation}\label{Gov1}
K_1(P,T)=\frac{a_{HCO_3^{-}}a_{H^{+}}}{a_{CO_2}a_{H_2O}}
\end{equation}
\begin{equation}\label{Gov2}
K_2(P,T)=\frac{a_{CO_3^{2-}}a_{H^{+}}}{a_{HCO_3^{-}}}
\end{equation} 
 
where $K_1$ and $K_2$ are referred to as the first and second dissociation (i.e. ionization) constants of carbonic acid, respectively. $a_i$ is the activity of the $i$-th chemical species. The activity of a chemical species is its concentration ($m_i$, using here the molality scale) times its activity coefficient ($\gamma_i$). In the Debye-H\"uckel theory, which usually serves as the theoretical foundation for describing solutions containing ionic solutes, a natural parameter describing the solution is its ionic strength \citep{Davidson2003statistical}:
\begin{equation}
I\equiv\frac{1}{2}\sum_im_iz_i^2
\end{equation}
where $z_i$ is the valence of the $i$-th ionic species. 

In order to estimate the activity of an ionic solute one requires an estimation for its activity coefficient. The activity coefficient is the fraction of the concentration of a component that is readily active in the solution. 
As long as the concentration of dissolved ions is not too high the main reason for an activity coefficient correction (i.e. correction for solution non-ideality) comes from electric shielding between unlike ions. The shielding effectively reduces the long range interaction rendering some portion of the ions less "active".  For electric shielding the most important aspect is the concentration of ions of a given valence rather then their elemental composition. Therefore, electric shielding can approximately be described with the help of the ionic strength parameter alone. 
In the Debye-H\"uckel theory only long range interactions between the different ions in the solution are taken into account. This implies the Debye-H\"uckel theory is a good approximation for rather dilute solutions. In this case the activity coefficient is theoretically given by the following form \citep[see page $507$ in][]{Davidson2003statistical}:
\begin{equation}\label{DaviesActivityCoeff}
\log_{10}\gamma_i=-1.8246\times 10^6 z_i^2 \frac{\left(\epsilon T\right)^{-3/2}\left(\rho_sI\right)^{1/2}}{1+50.29d\left(\epsilon T\right)^{-1/2}\left(\rho_sI\right)^{1/2}}
\end{equation}
where $\epsilon$ and $\rho_s$ are the dielectric constant and bulk density of the solvent, respectively. $T$ is the temperature and $d$ is the ionic diameter in angstroms. 
Since only long range interactions are considered in the Debye-H\"uckel theory, it is not applicable for solutions with high ionic strengths. It is found to be a good approximation as long as the ionic strength is less than $10^{-2.3}$\,molal \citep{zeebe2001}. 

\cite{Davies1938} suggested that a very simple empirical correction should be added to the Debye-H\"uckel activity coefficient, yielding what is known as the Davies activity coefficient equation: 
\begin{equation}\label{DaviesActivityCoeff}
\log_{10}\gamma_i=-1.8246\times 10^6\left(\epsilon T\right)^{-3/2} z_i^2\left(\frac{\sqrt{I}}{1+\sqrt{I}}-0.2I\right)
\end{equation}
Comparison with experimentally deduced equilibrium constants of various salt solutions in water, \cite{Davies1938} argued that eq.($\ref{DaviesActivityCoeff}$) has an accuracy of about $2\%$ for ionic strengths up to $0.1$\,molal. In practice the Davies activity coefficient is considered to be a good approximation for ionic strengths up to $0.5$\,molal \citep{zeebe2001}. It is important to notice that in the Davis equation the ionic diameter, $d$, is omitted and so ions of the same valence have equal activity coefficients. 

\cite{Ozsoy2008} measured the ionic composition of several rivers discharging into the Cilician basin. From their measurements at stations close to the rivers' sources we can derive an average ionic strength of $0.005$\,molal. Measurements therefore show that for freshwater on Earth the above equations for the activity coefficient are applicable.    

As one considers solutions of higher ionic concentrations two phenomena become increasingly important: ion pairing and ion complex formation. Ion pairing is when two ions of opposite charge are close enough to experience a strong coulombic interaction yet create no covalent bonds. An ion pair can be two ions that either are in contact and share a hydration shell, have a single solvent molecule in between them or have separate hydration shells \citep{Eva2013}. On the other hand, an ion complex forms when ions begin to share electrons, thus forming new ionic species. For example, Earth's seawater has on average an ionic strength of $0.7$\,molal \citep[see the standard mean chemical composition of seawater in appendix A in][]{zeebe2001}. Indeed, ion complexes between OH$^{-}$ or CO$_3^{2-}$ and divalent or trivalent metals are dominant in seawater in comparison to free metal ions \citep{Millero2009}. However, this is somewhat dependent on the metal under consideration. Therefore, in addition to their dependence on the ionic strength these higher ionic concentration phenomena depend on the type of element considered as well. Thus, activity coefficients that would adequately describe such phenomena could not depend on the ionic strength alone.
       
\cite{Guggenheim1935} argued that the Debye-H\"uckel theory can be made applicable to electrolyte solutions of higher ionic strengths. He suggested adding a correction accounting for short range interactions between neighbouring pairs of unlike-charged ions, i.e a second virial correction. This second virial correction would consist of adjustable parameters to be obtained by fitting to experimental data and would be dependent on the types of ions composing the existing pairs in the solution. The approach suggested by Guggenheim gave accurate activity coefficients, within experimental error, for ionic strengths up to $0.1$\,molal \citep{Guggenheim1955}. 
A model for the activity coefficients that is commonly used for describing seawater is that of \cite{Pitzer1973}.
In this model the constant coefficients of the second virial correction of \cite{Guggenheim1935} are taken to be functions of the ionic strength of the solution, and higher order correction terms are added to account for short range interactions between triplets of ions as well. The Pitzer model is extensively used when modelling electrolyte solutions where ionic concentrations exceed $1$\,molal. The parameters for the Pitzer model are usually obtained by a fit to solubility experimental data \citep[see][and references therein]{Li2011}. In this work we solve for the speciation of the carbonate system in the oceans of water-rich planets, and derive the resulting ionic strength. We then make sure our choice for the activity coefficient model is consistent with our results.

Before leaving the subject of the activity coefficient one may question the applicability of the Debye-H\"uckel theory to highly pressurized solutions. \cite{Manning2013c} estimated theoretically the solubility of corundum in KOH and aqueous solutions at $1$\,GPa, for various activity coefficient models, including the Davies equation. By comparing his predicted solubility to experimental data at this high pressure the author was able to show that the extended Debye-H\"uckel activity theory, and the Davies equation in specific, was adequate for modelling high pressure solubility. \cite{Manning2013b} repeated the same procedure, and compared his predicted solubility for calcite in water with experimental data at $1$\,GPa, where, once again the Davies equation proved adequate. We note that $1$\,GPa approximates the upper limit on the pressure expected at the bottom of our water planet oceans.

In the ocean, dissolved inorganic carbon resides in the species: CO$_2$, HCO$_3^{-}$ and CO$_3^{2-}$. The total concentration of the dissolved inorganic carbon is therefore \citep{zeebe2001}:
\begin{equation}\label{Gov3}
DIC \equiv \left[CO_2\right]+\left[HCO_3^{-}\right]+\left[CO_3^{2-}\right]
\end{equation}
where square brackets denote concentration throughout this work. 

Another governing equation that must be addressed is the charge neutrality requirement for the ocean. A variety of ions differing in their: valences, concentrations and elemental composition are dissolved in Earth's oceans. This ought also be the case for any ocean with a rocky floor. These introduce a higher level of complexity, because they may react with the carbonate system. For example, in the formation and solubility of calcite, or when ion rich silicates form clays during silicate weathering. 

When studying Earth's oceans one may partially overcome this enormous complexity by introducing the concept of carbonate alkalinity, defined as \citep{zeebe2001}:
\begin{equation}\label{Gov4}
CA\equiv \left[HCO_3^{-}\right]+2\left[CO_3^{2-}\right]
\end{equation}   
Eqs.($\ref{Gov1}$), ($\ref{Gov2}$), ($\ref{Gov3}$) and ($\ref{Gov4}$) represent four equations governing over six variables: $\left[CO_2\right]$,$\left[HCO_3^{-}\right]$, $\left[CO_3^{2-}\right]$, $\left[H^+\right]$, $DIC$ and $CA$. If one can measure any two of these variables the system can be solved for \citep{Millero2002}. This is indeed the practice for studying Earth's oceans, where the $DIC$ and the ocean's pH can be measured directly with relative ease \citep{zeebe2001}. No such short cuts are available for exoplanets whose oceans float on a rocky floor.  

For the case of Earth most of the carbon is deposited in rocks \citep{CarbonCycle}. Therefore, putting an upper limit on the allowed value for the $DIC$ (as expected for a closed system). Considering the pH for Earth's ocean, one can solve for the carbonate system speciation, and show that bicarbonate and carbonate are the dominant species in the $DIC$. In other words, the ocean makes use of the acidic nature of freely dissolved CO$_2$ to comply with the requirement for charge neutrality by transforming it to ionic species. 

It is interesting to note that, because the $DIC$ in Earth's oceans has an upper limit, our oceans can largely remain under-saturated with respect to freely dissolved CO$_2$. The consequence of which is the instability of SI CO$_2$ clathrate hydrate in the oceans on Earth. This may be a general conclusion for water worlds having a global ocean with a rocky bottom surface, if these oceans as well represent a closed system with respect to CO$_2$ (i.e. limited DIC). 
However, in this study, we focus on ocean planets whose ocean floor is composed of high pressure ice. Such planets can keep their oceans saturated in CO$_2$ and stabilize the appropriate clathrate hydrate \citep{Levi2017}. 

As discussed in the previous section the oceans we study here are likely poor in salts.   
Therefore, the charge neutrality of the ocean may be approximated as:
\begin{equation}\label{Gov4WaterPlanet}
0=\left[HCO_3^{-}\right]+2\left[CO_3^{2-}\right]+\left[OH^{-}\right]-\left[H^{+}\right]
\end{equation} 
Finally, when in equilibrium, the dissociation products of water obey:
\begin{equation}\label{Gov5}
K_w(P,T)=a_{H^{+}}a_{OH^{-}}
\end{equation}
Eqs.(\ref{Gov1}), (\ref{Gov2}), (\ref{Gov3}), (\ref{Gov4WaterPlanet}) and (\ref{Gov5}) govern the speciation of the carbonate system in our studied water planets. The six variables of the system are: $\left[CO_2\right]$,$\left[HCO_3^{-}\right]$, $\left[CO_3^{2-}\right]$, $\left[H^+\right]$, $\left[OH^-\right]$ and the $DIC$. Therefore, knowing one of the variables we can solve for the entire system. Because our studied oceans are saturated in CO$_2$, as mandated from being an open system, we can solve the system of governing equations as a function of $\left[CO_2\right]$.

\subsection{The Dissociation Constants}

In this subsection we discuss the first, $K_1$, and second, $K_2$, dissociation constants of carbonic acid, and the dissociation constant for water, $K_w$.
Experiments measuring $K_1$ and $K_2$ do not cover our full parameter space of interest. 
For example, \cite{Li2007} represented $K_1$ and $K_2$ dependence on the temperature and pressure using a polynomial equation. The polynomial coefficients, for the part representing the temperature dependence, were obtained by a fit to $K_1$ and $K_2$ experimental data. For the dissociation constants' dependency on the pressure the authors used experimental partial molar volumes and compressibility data. This procedure is reported to reproduce the available experimental data with good accuracy in the temperature range from $0^\circ$C to $250^\circ$C. This temperature range is adequate for the purpose of modelling water-rich planets. However, the applicability of the formulations of \cite{Li2007} have an upper limit on the pressure of $0.1$\,GPa. This upper limit is sufficient for modelling the carbonate system at the bottom of Earth's ocean. However, it is an order of magnitude lower than the pressure prevailing at the ocean floor, for our studied water planets. 
 
Experiments determining the equilibrium ionization constants at low temperatures (approximately $300$\,K) and for pressures exceeding $0.1$\,GPa are very few in number \citep[see review of experiments in][]{Li2007}. The notable exceptions are the experiments of \cite{Ellis1959} reaching up to $0.3$\,GPa and \cite{Read1975} reaching as high as $0.2$\,GPa, and both are only for $K_1$.

High pressure ($\geq 1$\,GPa) carbonic acid ionization constants are experimentally available. However, the parameter space probed by experiments is mostly guided by the conditions within planet Earth. At these relatively high pressures, and within the Earth, one examines the deep crust or upper mantle. Because the thermal conductive profile rapidly increases the temperature with depth in Earth's crust, then at $1$\,GPa, temperatures fall in the range of $700$\,K-$1300$\,K, depending on the type of tectonic environment \citep{Jones1983}. Therefore, efforts for evaluating $K_1$ and $K_2$ at high pressure were so far focused on temperatures exceeding, by at least a factor of two, those that are of interest to us. 

A better understanding of the carbon cycle on Earth requires quantifying the dissolution of calcite and aragonite in aqueous fluids that are released from the tectonic slab during its subduction. We note that, \cite{Facq2014} hydro-statically compressed a sample of calcite immersed in double-distilled water using a heated diamond anvil cell at temperatures appropriate for describing cold subduction zones ($573$\,K-$673$\,K) to deep-crust pressures as high as $8$\,GPa.   

In this subsection we will make use of two very common methods for interpolating between and extrapolating over the experimentally determined ionization constants: the solvent density method, and the revised Helgeson-Kirkham-Flowers (HKF) equation of state for electrolyte solutions \citep[see review of methods in][]{Dolejs2013}. Our approach will be to estimate the free parameters for each of the two methods by fitting to low pressure ($\sim 0.1$\,GPa) experimental data for the dissociation constants. Then we use models to extrapolate to high pressure. We then test the extrapolations by their ability to predict the high pressure experimental dissociation constants given in \cite{Facq2014}. The perils of extrapolating beyond the experimental data without a sound theoretical model will also be shown and discussed in this subsection.  

First, we examine the solvent density method. We note that it has several variants \citep[see discussion in][]{Dolejs2013}. The physical reasoning at the base of this method is: the solvent (water) can be described as a continuous dielectric medium, that the free energy change of solvation of an ion in the solvent can be described to a good approximation with the aid of the solvent's dielectric constant, and that this dielectric constant is strongly correlated to the solvent's density. 
As an example we adopt the formulation given in \cite{Dolejs2010}:
\begin{equation}\label{KDensityModel}
\ln K = \frac{A_1}{T}+A_2+A_3\ln T+A_4T+A_5\ln\rho_w \equiv m(T)+A_5\ln\rho_w
\end{equation}  
where $K$ is the ionization constant, $\rho_w$ is the density of water and $T$ is the temperature. The form for $m(T)$ comes from assuming that the neutral molecule breakdown followed by the hydration of the products can be modelled using a free energy derived from integration over heat capacity terms, which are assumed to be linear functions of the temperature. The term $\ln\rho_w$ is assumed to encapsulate the effects of electrostriction. Plotting the experimental isothermal data from \cite{Read1975} and \cite{Ellis1959} as $\ln K_1$ versus $\ln\rho_w$ we indeed find it to be linear to a good approximation. We also find that $A_5$ must be varied with the temperature, in order to fit the data. Adopting the parametrization for the dielectric constant of water from \cite{Sverjensky2014}, we assign to $A_5$ the following form:
\begin{equation}\label{a5fit}
A_5(T)=b_1T+b_2\sqrt{T}+b_3
\end{equation}
For each isothermal data set reported in \cite{Read1975} and \cite{Ellis1959} we obtain a value for $m$ and $A_5$ using a linear fit to the data, thus obtaining $m(T)$ and $A_5(T)$. Then, using the least squares method we obtain the optimal values for the parameters $A_1$ through $A_4$ and $b_1$ through $b_3$. The values we obtain for the coefficients $A_1$ to $A_4$ match the discrete experimentally fitted values for $m(T)$ with an absolute average deviation of $0.16$\%. The values we find for $b_1$ to $b_3$ produce the experimentally fitted values for $A_5(T)$ with an absolute average deviation of $3.2$\%.

\begin{figure}[ht]
\centering
\mbox{\subfigure{\includegraphics[width=7cm]{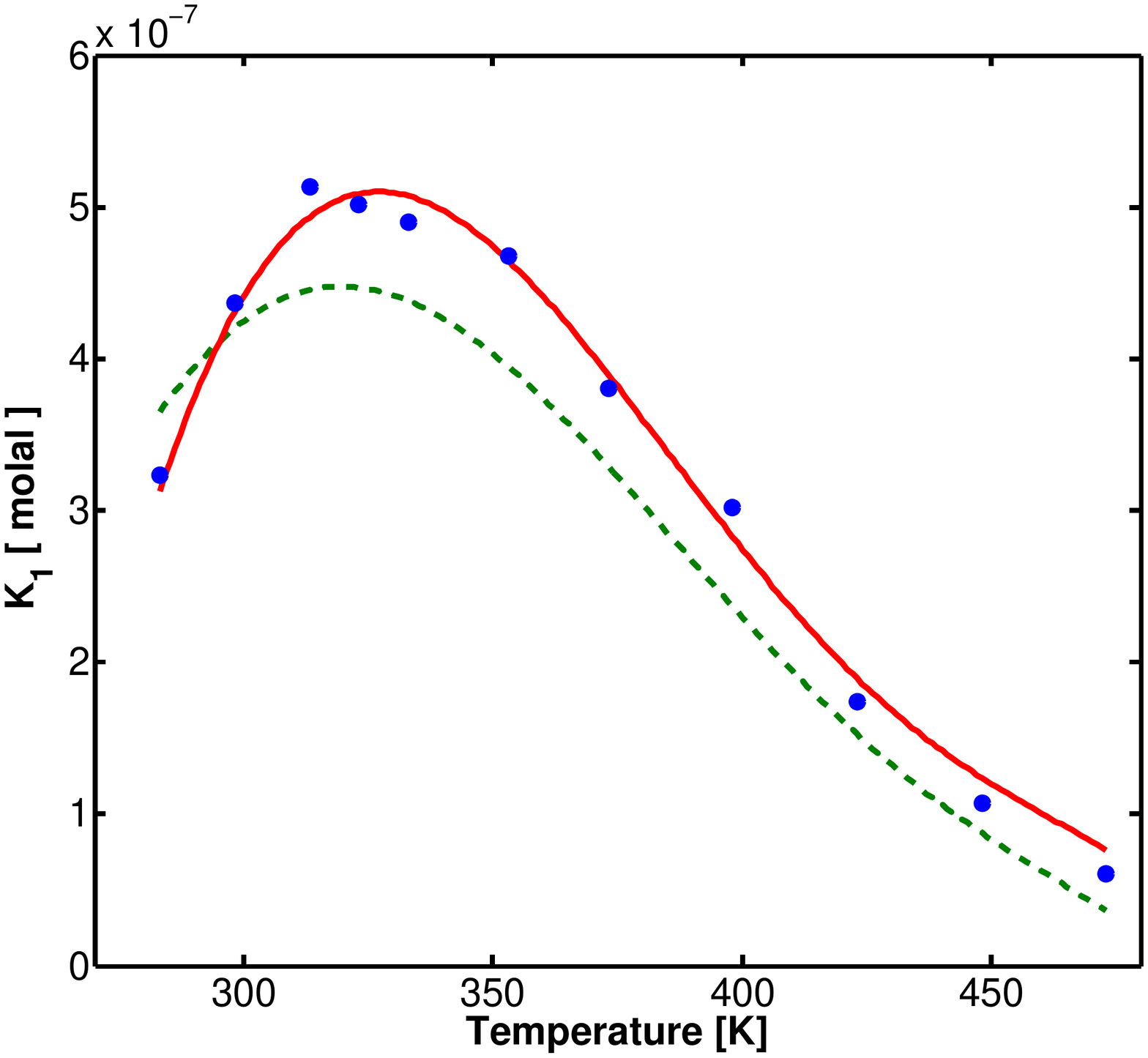}}\quad \subfigure{\includegraphics[width=7cm]{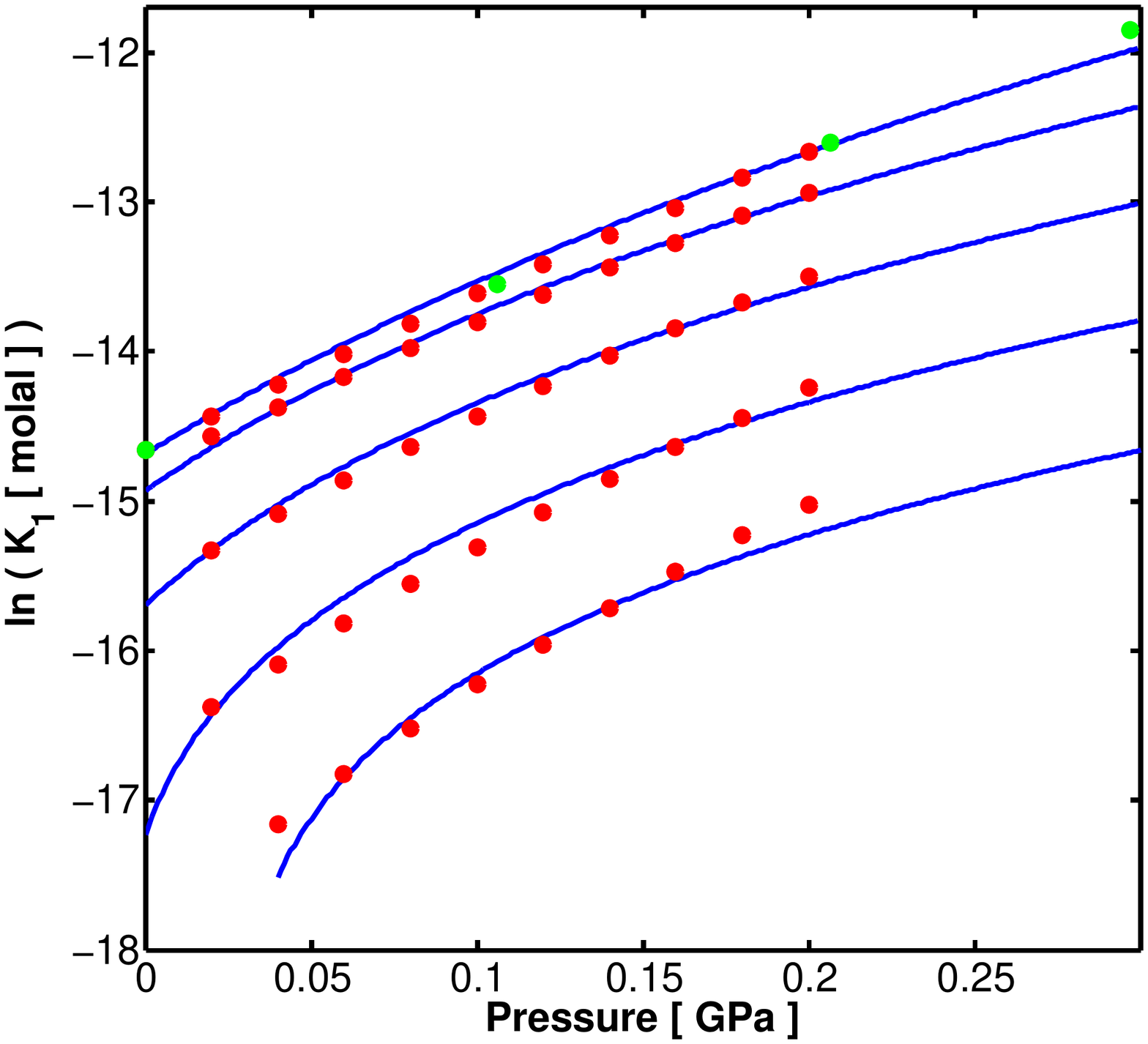}}}
\caption{\footnotesize{\textbf{Right panel} - the first ionization constant, $K_1$, versus pressure for five example isotherms: $25^\circ$C, $100^\circ$C, $150^\circ$C, $200.1^\circ$C, $250.1^\circ$C, where the lower temperature is higher in the figure. Red and green filled circles are data points from \cite{Read1975} and \cite{Ellis1959} respectively. Solid blue curves are the density model fit of Eqs.(\ref{KDensityModel}) and (\ref{a5fit}). \textbf{Left panel} - the first ionization constant, $K_1$, along the water saturation vapour pressure curve. Filled blue circles are data points from \cite{Stefansson2013}. The dashed green curve is from the solvent density model of Eqs.(\ref{KDensityModel}) and (\ref{a5fit}). The solid red curve is from the revised-HKF model.}}
\label{fig:fitdatadensity}
\end{figure}

The formulation for $K_1$ given in Eqs.(\ref{KDensityModel}) and (\ref{a5fit}), with the fitted parameters, reproduces the data sets of \cite{Read1975} and \cite{Ellis1959} with an absolute average deviation of a few percent, reaching as high as $13$\% for the case of the $200.5^\circ$C isotherm from \cite{Read1975} (see right panel in Fig.\ref{fig:fitdatadensity}). We have also tried to simultaneously reproduce the experimental values for $K_1$ along the water saturation vapour pressure curve from \cite{Stefansson2013}. Our solvent density model fit can reproduce the latter experimental data set with an absolute average deviation of $15.6$\% (see left panel in Fig.\ref{fig:fitdatadensity}).

Our results show that the solvent density model can provide a simple method for interpolating between data points to an accuracy of a few percent. Considering its simplicity it is a valuable tool. In addition, we test this model as an extrapolation tool, by comparing its predictions to available experimental data. In table \ref{tab:HKFVDensityExtrapolation} we list the high temperature and high pressure experimental data from \cite{Facq2014} along side extrapolated $K_1$ values using the solvent density model, for the same temperature and pressure conditions. At a pressure of $1$\,GPa the extrapolated values, using the solvent density model, are good to a factor of a few. However, at higher pressures there are orders of magnitude deviations. Therefore, using this theoretical model to extrapolate on the experimental data would produce erroneous results.

\begin{deluxetable}{cccccccc}
\tablecolumns{8}
\tablewidth{0pc}
\tablecaption{A Comparison Between Extrapolated and Experimental Data }
\tablehead{
\colhead{} & \colhead{}  & \colhead{$1$} & \colhead{$2$}  & \colhead{$3$}  & \colhead{$4$} & \colhead{$5$} & \colhead{$6$}  \\
\colhead{} & \colhead{} & \colhead{[GPa]} & \colhead{[GPa]} & \colhead{[GPa]} & \colhead{[GPa]} & \colhead{[GPa]}  &  \colhead{[GPa]} }
\startdata

\multirow{3}{*}{$300^\circ$C ($K_1$)}  & Exp.    & -5.38 & -3.64  & -2.13  & -0.75 &  0.54  &   \\
                                       & Density & -5.97 & -5.48  & -5.18  & -4.96 & -4.79  &  - \\
                                       & HKF     & -5.33 & -3.75  & -2.45  & -1.30 & -0.26  &   \\

\hline
\multirow{3}{*}{$350^\circ$C ($K_1$)}  & Exp.    & -5.90 & -4.23  & -2.81  & -1.52 & -0.31  & 0.84  \\
                                       & Density & -6.36 & -5.85  & -5.55  & -5.33 & -5.16  & -5.02  \\
                                       & HKF     & -5.80 & -4.26  & -3.02  & -1.93 & -0.94  & -0.02   \\

\hline
\multirow{3}{*}{$400^\circ$C ($K_1$)}  & Exp.    & -6.41 & -4.80  & -3.45  & -2.23 & -1.09  & -0.02  \\
                                       & Density & -6.74 & -6.19  & -5.86  & -5.63 & -5.45  & -5.30  \\
                                       & HKF     & -6.25 & -4.76  & -3.56  & -2.52 & -1.57  & -0.70   \\

\hline\hline

\multirow{2}{*}{$300^\circ$C ($K_2$)}  & Exp.    & -8.30 & -7.10  & -6.3   & -5.2  &  -4.4  & -  \\
                                       & HKF     & -8.53 & -7.25  & -6.19  & -5.22 &  -4.32 &    \\

\hline
\multirow{2}{*}{$350^\circ$C ($K_2$)}  & Exp.    & -8.38 & -7.18  & -6.3   & -5.3  & -4.6   & -3.8  \\
                                       & HKF     & -8.64 & -7.36  & -6.33  & -5.41 & -4.54  & -3.72   \\

\hline
\multirow{2}{*}{$400^\circ$C ($K_2$)}  & Exp.    & -8.46 & -7.26  & -6.3   & -5.5  & -4.7   & -4.0  \\
                                       & HKF     & -8.76 & -7.46  & -6.45  & -5.55 & -4.72  & -3.94   \\

\enddata
\tablecomments{\footnotesize{A comparison between experimental values for $K_1$ and $K_2$ \citep{Facq2014} and values obtained by extrapolation using the solvent density model (denoted here as Density) and the revised-HKF model (denoted here as HKF). Tabulated values are log base 10 of the first and second ionization constants of carbonic acid.}}
\label{tab:HKFVDensityExtrapolation}
\end{deluxetable}

A different theory is the Helgeson-Kirkham-Flowers (HKF) model. A neutral molecule dissolved in a medium can react with its surrounding molecules, and form ions. For example, a CO$_2$ molecule in an aqueous fluid can react with water molecules and form bicarbonate. The result is an increased electrostatic field in the locality surrounding the ionic product. The increased electrostatic field deforms the surrounding dielectric solvent, a phenomenon termed "electrostriction" \citep{stratton2007electromagnetic}. 
Electrostriction is actually a sum of many interdependent phenomena, that involve the ion-solvent interaction and structural deformation of the solvent shells surrounding the ion. \cite{Helgeson1976} proposed a simplified way to describe the thermodynamic properties of electrolyte aqueous solutions. Their simplified model for describing electrostriction is the foundation of the widely used revised-HKF model. 

\cite{Helgeson1976} argued, that every thermodynamic quantity, of dissolved ions in a dielectric solvent, can be broken down to a sum of three terms. Those are: an ionic intrinsic term, a solvent structural term, and a direct ion-solvent interaction term. The last two terms combined represent the net effect of electrostriction. Taking the standard partial molal volume as an example for such a thermodynamic quantity, its value may be written down as:
\begin{equation}
\bar{V}^\circ = \bar{V}^\circ_i + \Delta\bar{V}^\circ_e = \bar{V}^\circ_i + \Delta\bar{V}^\circ_c + \Delta\bar{V}^\circ_s
\end{equation}   
Here, $\bar{V}^\circ_i$ represents an intrinsic ionic volume independent of the nature of the solvent. $\Delta\bar{V}^\circ_e$ is the volume change associated with electrostriction, which is further broken down to: a volume change from structural collapse in the solvent surrounding the dissolved ion, $\Delta\bar{V}^\circ_c$, and a volume change associated with direct ion-solvent interactions, also called volume change of solvation, $\Delta\bar{V}^\circ_s$.

Consider a neutral conducting sphere of radius $\hat{b}$, that is immersed in a medium with a dielectric constant $\epsilon$. The work required to charge it to an ionic valence $z$ is \citep{Davidson2003statistical}:
\begin{equation}
W=\frac{1}{2}\frac{z^2q^2}{\epsilon \hat{b}}
\end{equation} 
where $q$ is the magnitude of fundamental charge.  The free energy change associated with transferring such a charged sphere from a vacuum to a dielectric medium of dielectric constant $\epsilon$ is therefore:
\begin{equation}
\Delta \bar{G}^\circ_s = \frac{1}{2}\frac{z^2q^2}{\hat{b}}\left( \frac{1}{\epsilon} -1 \right)\equiv\omega\left( \frac{1}{\epsilon} -1 \right)
\end{equation}
where $\omega$ is the Born coefficient.
This last function is also known as the Born equation, and was used in \cite{Helgeson1976} to quantify the standard partial molal Gibbs free energy of solvation. Within the framework of the HKF model, derivatives of this formulation produce, for every thermodynamic quantity, the solvation contribution to the net electrostriction effect. This is in contradiction with other theoretical models where the last term is considered to represent the effect of electrostriction in its entirety \citep[e.g.][]{Benson1963}. 
Thus, the standard partial molal volume change of solvation, the standard partial molal compressibility of solvation and standard partial molal isobaric heat capacity of solvation have the following forms \citep{Helgeson1976,Helgeson1981}:
\begin{eqnarray}
\Delta\bar{V}^\circ_s = -\frac{\omega}{\epsilon}\left(\frac{\partial\ln{\epsilon}}{\partial P}\right)_T  \\
\bar{\kappa}^\circ_s \equiv -\left(\frac{\partial\Delta\bar{V}^\circ_s }{\partial P}\right)_T =  \frac{\omega}{\epsilon}\left[\left(\frac{\partial^2\ln{\epsilon}}{\partial P^2}\right)_T - \left(\frac{\partial\ln{\epsilon}}{\partial P}\right)^2_T\right]   \\
\bar{C}^\circ_{p,s} = \frac{\omega T}{\epsilon^2}\left[\left(\frac{\partial^2\epsilon}{\partial T^2}\right)_P-\frac{2}{\epsilon}\left(\frac{\partial\epsilon}{\partial T}\right)_P^2\right]
\end{eqnarray}

\cite{Helgeson1976} examined the intrinsic and solvent structural collapse effects by subtracting from experimental values for the ionic standard partial molal volumes the theoretical solvation volume change effect. Repeating this algorithm with experimental standard partial molal compressibility data they have formulated their results in the following form:
 \begin{equation}\label{IntStrucVolume}
\bar{V}^\circ_i + \Delta\bar{V}^\circ_c = a_1+a_2P+a_3\frac{T}{T-\Theta}+a_4P\frac{T}{T-\Theta}
\end{equation}
\cite{Helgeson1981} further subtracted from experimental data for the standard partial molal isobaric heat capacities of electrolytes, at water saturation vapour conditions, the theoretical contribution from solvation. They then found for the intrinsic and solvent structural collapse contributions, at $1$\,bar, the following:
\begin{equation}\label{IntStrucHeatCapacity}
\bar{C}^\circ_{p,i} + \bar{C}^\circ_{p,c} = c_1+c_2\frac{1}{T-\Theta}
\end{equation} 
The nature of the asymptotic behaviour at the temperature $\Theta$ is still under debate. Although the asymptotic behaviour in the HKF model was arrived at by a simple fit to experimental data, there seems to be a real physical basis for this behaviour. Several of the thermodynamic properties of water seem to diverge with decreasing temperature in the supercooled liquid regime, with a singularity in the temperature falling in the range from $223$\,K to $228$\,K \citep{Torre2004}. The nature of this behaviour could be due to polymorphism of liquid water, where the singularity in the temperature could be indicative of a transition between a high density and a low density liquid \citep{Mallamace2014}. Another debated possibility is that the singularity in the temperature represents a state of dynamic self-arrest. Relaxation times have a Arrhenius type behaviour with some activation energy. For the case of water the activation energy increases drastically with decreasing temperature in the supercooled regime, and seems to turn infinite at some finite temperature, $\Theta$, resulting in dynamic arrest \citep{Hecksher2008}. In the HKF model a constant value of $228$\,K is adopted for $\Theta$. 

\cite{Tanger1988} introduced two corrections to the formulations given above. They have formulated what is known as the revised-HKF equations of state. The asymptotic behaviour may have a different exponent for different thermodynamic properties. On the grounds of a better fit to heat capacity experimental data \cite{Tanger1988} replaced Eq.($\ref{IntStrucHeatCapacity}$) with the following:
\begin{equation}
\bar{C}^\circ_{p,i} + \bar{C}^\circ_{p,c} = c_1+c_2\left(\frac{1}{T-\Theta}\right)^2
\end{equation}
Eq.($\ref{IntStrucVolume}$) was found to predict molal volumes that approximate well to experimental data only up to $0.1$\,GPa. Therefore, using experimental data for the molal volumes of NaCl and K$_2$SO$_4$ to $1$\,GPa \cite{Tanger1988} revised Eq.($\ref{IntStrucVolume}$) to the following form:
\begin{equation}
\bar{V}^\circ_i + \Delta\bar{V}^\circ_c = a_1+a_2\frac{1}{\Psi + P}+a_3\frac{1}{T-\Theta}+a_4\frac{1}{T-\Theta}\frac{1}{\Psi + P}
\end{equation}
where $\Psi$ is a constant depending on the type of solvent. 
Hence, the standard partial molal Gibbs free energy of formation of an aqueous species, either ion or electrolyte, may be derived, yielding the following form:
$$
\Delta\bar{G}^\circ_f = \Delta\bar{G}^\circ_{f,r}  - \bar{S}^\circ_r(T-T_r)-c_1\left[T\ln\frac{T}{T_r}-T+T_r\right] + a_1(P-P_r) 
$$
$$ 
+ a_2\ln\left(\frac{\Psi + P}{\Psi + P_r}\right) - c_2\left[\left\lbrace \frac{1}{T-\Theta}-\frac{1}{T_r-\Theta}\right\rbrace \frac{\Theta-T}{\Theta}-\frac{T}{\Theta^2}\ln\left(\frac{T_r(T-\Theta)}{T(T_r-\Theta)}\right)\right]
$$
$$
+\frac{1}{T-\Theta}\left[ a_3(P-P_r)+a_4\ln\left(\frac{\Psi + P}{\Psi + P_r}\right)\right] + \omega\left(\frac{1}{\epsilon}-1\right) - \omega_r\left(\frac{1}{\epsilon_r}-1\right)
$$
\begin{equation}
+\omega_r \frac{1}{\epsilon^2_r}\left(\frac{\partial\epsilon}{\partial T}\right)_{P=P_r}(T-T_r)
\end{equation} 
where the subindex, $r$, refers the parameter to its value at the reference conditions: $P_r=1$\,bar and $T_r=298.15$\,K. With the aid of the last equation one may derive the standard partial molal Gibbs free energy of reaction, for the first and second ionization constants of carbonic acid. This Gibbs free energy is related to the equilibrium constants $K_1$ and $K_2$ in the following way:
\begin{equation}
\ln K_1(P,T) = -\frac{1}{RT}\left( \Delta\bar{G}^\circ_{f,HCO_3^-} + \Delta\bar{G}^\circ_{f,H^+} - \Delta\bar{G}^\circ_{f,CO_2} - \Delta\bar{G}^\circ_{f,H_2O}\right) 
\end{equation}
\begin{equation}
\ln K_2(P,T) = -\frac{1}{RT}\left( \Delta\bar{G}^\circ_{f,CO_3^{-2}} + \Delta\bar{G}^\circ_{f,H^+} - \Delta\bar{G}^\circ_{f,HCO_3^-} \right)
\end{equation} 
where $R$ is the ideal gas constant per mole. Usually the apparent Gibbs free energies of formation are defined with respect to the value for the hydrogen ion. Hence, the actual apparent standard partial Gibbs free energy of formation for $H^+$ is by definition zero. 

The revised-HKF model has become widely used in geochemical calculations. The parameters for this model, for a plethora of ions and electrolytes, are given in the literature \citep[e.g.][]{Shock1988}. Also, correlation techniques were developed, that help estimate the parameters needed for this model, even for materials for which relatively little experimental data exists \citep{Shock1988}. 

\begin{figure}[ht]
\centering
\includegraphics[trim=0.2cm 4.3cm 0.2cm 5cm , scale=0.55, clip]{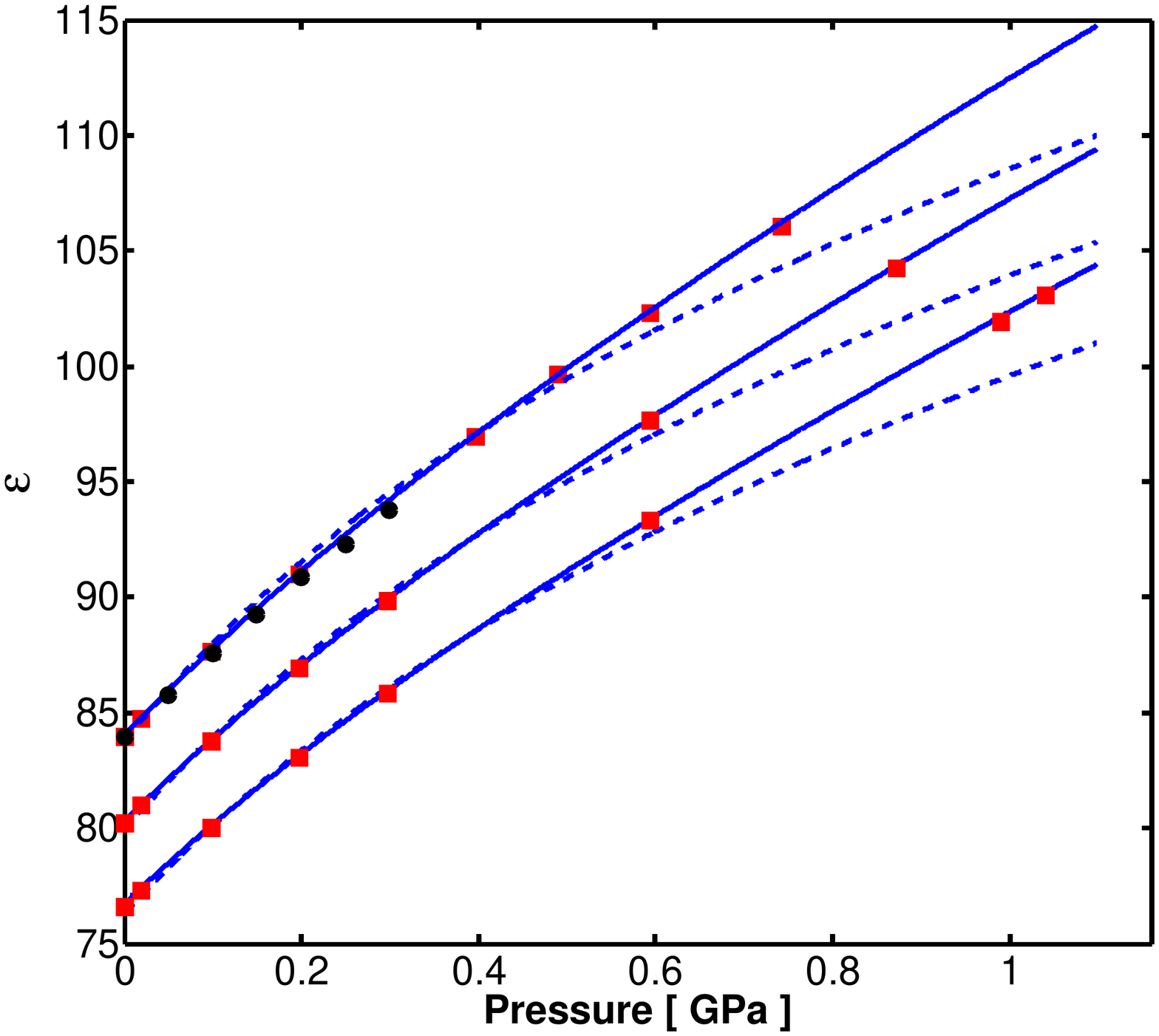}
\caption{\footnotesize{The dielectric constant of water versus pressure for three isotherms: $283$\,K, $293$\,K and $303$\,K. The dielectric constant decreases with increasing temperature for a given pressure. Solid blue curves are from the model of \cite{Fernandez1997}. Dashed blue curves are from the model of \cite{Marshall2008}. Filled black circles are data points from K.R. Srinivansan's dissertation \citep[see][for tabulated data]{Fernandez1995} and filled red squares are data points from the dissertation of W. L. Lees \citep[see][for tabulated data]{Fernandez1995}.}}
\label{fig:DielectricConstant}
\end{figure}

Investigating the possible ionic composition of aqueous fluids in Earth's deep-crust and mantle, with the aid of the revised-HKF model, requires the dielectric constant for water at the appropriate conditions \citep[see discussion in][]{Manning2013}. Therefore, most of the effort towards determining the value for the dielectric constant is concentrated on pressures and temperatures far exceeding those we are interested in \citep[e.g][]{Sverjensky2014}. Here, we examine two published fitted models for the dielectric constant of water: that of \cite{Marshall2008} and that of \cite{Fernandez1997}. \cite{Fernandez1995} tabulated most of the then available experimental data for the dielectric constant of water, which he then used to build a parametrized model \citep{Fernandez1997}. The latter was approved for release by the IAPWS (latest version Sept. 1997).          
In Fig.\ref{fig:DielectricConstant} we compare between the models of \cite{Marshall2008} and \cite{Fernandez1997}, by examining their ability to describe known experimental data. The model of \cite{Fernandez1997} is a better fit, and is adopted in this study.
We note that the density for liquid water used in this work is from \cite{wagner02}.
 
If one adopts the revised-HKF model parameters given in the literature, for the different ions and electrolytes, one also has to adopt all other formulations that were used to produce these values. This could present a problem if some of the original formulations become outdated. For example, we have used a more recent formulation for the Gibbs free energy of water \citep{wagner02}, rather then the one used to constrain many of the tabulated revised-HKF parameters \citep[see][and references therein]{Oelkers1995}. Another common practice is to assign neutral molecules (e.g. CO$_2$) with a non-zero Born coefficient, to aid in fitting the experimental data. This practice is in clear contradiction with the definition of the Born coefficient. We have decided not to adopt this approach. We believe that keeping the integrity of the physical foundation of the HKF model is an important prerequisite in assessing its ability to predict experimental data. Our approach required making small corrections to the published revised-HKF parameters, so that the model could fit the experimental dissociation constants of \cite{Ellis1959} and \cite{Read1975}. In table \ref{tab:revisedHKFparameters} we list the revised-HKF parameters used in this study.

The revised-HKF model fits the experimental data for K1, along the water saturation vapour pressure curve, \citep[data from][]{Stefansson2013} with an absolute average deviation of $6.9$\% (see left panel in Fig.\ref{fig:fitdatadensity}). The isothermal data sets from \cite{Ellis1959} and \cite{Read1975} for temperatures less than $200^\circ$C are reproduced with an absolute average deviation of a few percent. For the higher temperatures larger deviations occur reaching as high as $30$\% for the $250^\circ$C isotherm reported in \cite{Read1975}. This latter deviation may be further reduced, by adopting various models for the Born coefficient's dependence on temperature and pressure, and indeed this is often the course of action adopted. However, we have adopted a constant Born coefficient. This is because models for the Born coefficient dependence on pressure and temperature are often constrained by a fit to experimental data. A Born coefficient constrained in this way is appropriate only for the fitted ion and is likely non-transferable.

\begin{deluxetable}{lccc}
\tablecolumns{4}
\tablewidth{0pc}
\tablecaption{ Revised-HKF Model Parameters }
\tablehead{
\colhead{Revised-HKF Parameter } & \colhead{CO$_2$}  & \colhead{HCO$_3^-$} & \colhead{CO$_3^{-2}$}   \\
\colhead{} & \colhead{} & \colhead{} & \colhead{} }
\startdata

$\Delta\bar{G}^\circ_{f,r}$ [cal\,mol$^{-1}$]  &   -91881     &  -142110    &   -128010    \\
$\bar{S}^\circ_r$ [cal\,mol$^{-1}$\,K$^{-1}$]  &   27.2570    &   22.5600   &   -12.3025   \\
$a_1$ [cal\,mol$^{-1}$\,bar$^{-1}$]            &   0.6563     &   0.7409    &    0.5742    \\
$a_2$ [cal\,mol$^{-1}$]                        &   747.00     &   111.55    &      500     \\
$a_3$ [cal\,K\,mol$^{-1}$\,bar$^{-1}$]         &    2.81      &    1.23     &     -2.0     \\
$a_4$ [cal\,K\,mol$^{-1}$]                     & -30900       &   -28300    &   -108000    \\
$c_1$ [cal\,mol$^{-1}$\,K$^{-1}$]              &    33.600    &   18.116    &    27.1800   \\
$c_2$ [cal\,K\,mol$^{-1}$]                     &   237600     &  -261800    &   -600000    \\ 
$\omega_r$  [cal\,mol$^{-1}$]                  &      0       &   120960    &    460000    \\

\enddata
\tablecomments{\footnotesize{List of parameter values adopted in this study to be used with the revised-HKF model.}}
\label{tab:revisedHKFparameters}
\end{deluxetable}

For purposes of interpolating between experimental data points both the solvent density model and the revised-HKF model present similar accuracies. Since solvent density models are much simpler they could prove preferable. However, the test we are more interested in is that of extrapolation. Once again, we extrapolate, now using the revised-HKF model, whose parameters were fitted to the data of \cite{Ellis1959} and \cite{Read1975}. As before we compare these extrapolations with the high pressure data reported in \cite{Facq2014}. We list the results in table \ref{tab:HKFVDensityExtrapolation}. One can see that for pressures below $2$\,GPa the experimental data from \cite{Facq2014} differs from the extrapolated values by a multiplicative factor of no more than $1.45$. Therefore, we suggest an error of about $50$\% is adequate for our modelled values for $K_1$, at the pressures prevailing at the bottom of water planet oceans.
 
Even for the highest experimental pressures reported in \cite{Facq2014}, our extrapolated values using the revised-HKF model never differ from the experimental data by more than a factor of a few. Therefore, when it comes to extrapolation the revised-HKF model is more reliable than the solvent density model, and is therefore used in this study to model both $K_1$ and $K_2$. 

There is even less reported experimental data, at high pressure and low temperature, for the second ionization constant, $K_2$, than there is for $K_1$ \citep[see table $1$ in][]{Li2007}.  
In order to extrapolate the experimental data for $K_2$ up to approximately $1$\,GPa, for our desired low temperatures, we again use the revised-HKF model. We constrain the revised-HKF model parameters by fitting to experimental data for $K_2$: along the saturation vapour pressure curve \citep{Stefansson2013}, and room temperature data up to $0.1$\,GPa \citep{Hershey1983}. The adopted parameters are listed in table \ref{tab:revisedHKFparameters}. With the aid of these adjusted revised-HKF parameters we extrapolate to high pressure and temperature, and compare with the measurements of \cite{Facq2014} for $K_2$. Here as well, we adopt the disagreement between the two as our probable error.

\begin{figure}[ht]
\centering
\includegraphics[trim=0.2cm 4.3cm 0.2cm 4.5cm , scale=0.55, clip]{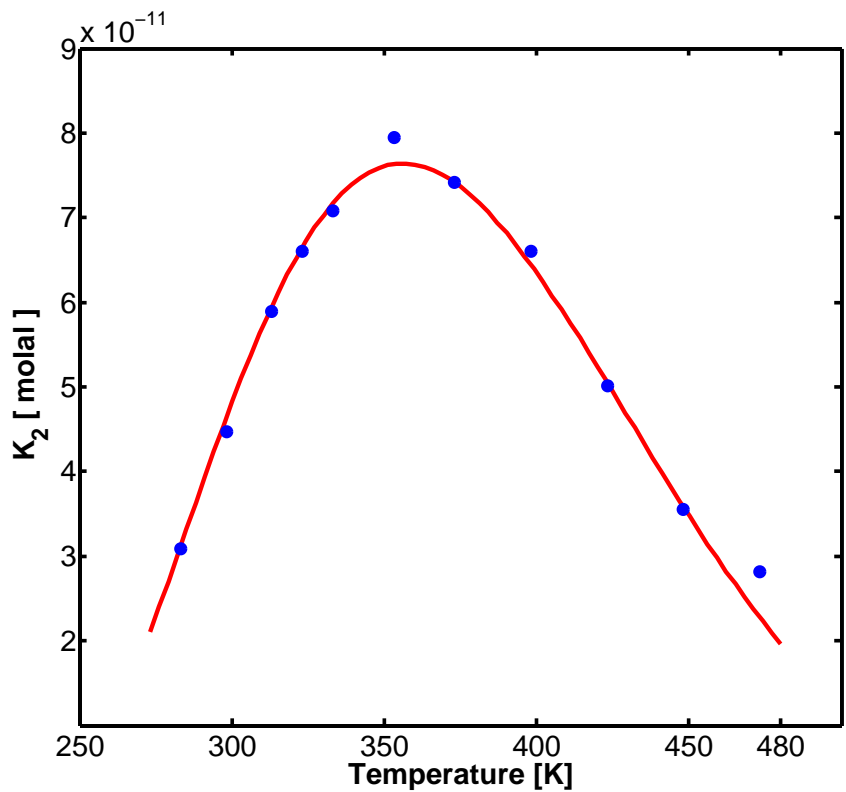}
\caption{\footnotesize{The second dissociation constant of carbonic acid, $K_2$, along the water saturation vapour pressure curve. Solid red curve is the fitted revised-HKF model with parameters tabulated in table \ref{tab:revisedHKFparameters}. Filled blue circles are experimental data points from \cite{Stefansson2013}.}}
\label{fig:K2VapPressure}
\end{figure}

The revised-HKF model, with our adopted parameters, fits the experimental data for $K_2$, along the water saturation vapour pressure curve \citep{Stefansson2013} with an absolute average deviation of $3.18$\% (see Fig.\ref{fig:K2VapPressure}). The absolute average deviation between our fitted revised-HKF model and the data from \cite{Hershey1983} is $0.95$\%. In table \ref{tab:HKFVDensityExtrapolation} we list the experimental values from \cite{Facq2014} for $K_2$, together with the values obtained by extrapolation using the revised-HKF model. When averaged over the entire experimental pressure range tested for in \cite{Facq2014} the disagreement between extrapolated and experimental values is approximately $25$\%. At $1$\,GPa, approximately the ocean bottom pressure, the revised-HKF model predicts values for $K_2$ which are as much as a factor of $2$ smaller than the experimental data. This factor of $2$ will be our assumed error, and we will test for its significance on the carbonate speciation and ocean pH below. 

The dissociation constant of water, $K_w$, is taken from \cite{Bandura2006}. This model has been adopted by the IAPWS (publication Aug. 2007). These authors have modelled the Gibbs free energy of the water dissociation reaction by separating it to an ideal gas and a residual contribution. The residual contribution was estimated by a sum of three terms: The work required to form proper cavities in the water bulk, the ion-water potential of interaction, and the effect of the polarization of the water bulk. The latter was modelled with the aid of the Born equation. The free parameters of their model were obtained by a fit to experimental data. The resulting model is able to reproduce the experimental data, with high accuracy, in the temperature range of $0^\circ$C-$800^\circ$C, and pressures up to $1$\,GPa. This $P-T$ range covers the conditions of interest to us.

\section{OCEAN pH and CARBONATE SPECIATION}

In this section we present the numerical results for the system described above. 
We approximate the thermal profile in the ocean with an isotherm. We solve for three cases: $275$\,K, $280$\,K and $285$\,K. These values are reasonable because, our studied planets have a thick ice mantle which is deprived of radioactive components. Therefore, heat fluxes at the ocean floor are probably low \citep[see][for heat flux approximations]{Levi2014}. For these temperatures the solubility of CO$_2$ in the bulk of the ocean is governed by the equilibrium with SI clathrate hydrates of CO$_2$ \citep{Levi2017}. 

For convenience, we have cast into a cubic polynomial form the first and second dissociation constants of carbonic acid, as derived from the revised-HKF equation of state:          
\begin{equation}\label{PolyAssist}
\log_{10}{K} = \alpha_1 P^3 + \alpha_2 P^2 + \alpha_3 P +\alpha_4
\end{equation} 
where $P$ is in GPa and $K$ is in molal. For each isotherm, the polynomial is valid for the pressure range spanning from the surface of the ocean to the ocean floor. The coefficients for the polynomials are tabulated in table \ref{tab:K1K2PolynomialCoeff}. 

\begin{deluxetable}{cccccccc}
\tablecolumns{8}
\tablewidth{0pc}
\tablecaption{Polynomial coefficients to be used with Eq.($\ref{PolyAssist}$). }
\tablehead{
\colhead{$T$} & \colhead{K} & \colhead{$\alpha_1$} & \colhead{$\alpha_2$} & \colhead{$\alpha_3$} & \colhead{$\alpha_4$}  \\
\colhead{[K]} & \colhead{[molal]} & \colhead{} & \colhead{} & \colhead{} & \colhead{} }
\startdata
\multirow{2}{*}{$275$} & $K_1$ & 1.6673 & -2.9143  & 4.6803  & -6.634  \\
                       & $K_2$ & 2.7847 & -5.0565  & 6.8889  & -10.629 \\
\hline
\multirow{2}{*}{$280$} & $K_1$ & 1.4898 & -2.8122  & 4.6586  & -6.5458 \\
                       & $K_2$ & 2.1813 & -4.2479  & 6.2391  & -10.542 \\
\hline
\multirow{2}{*}{$285$} & $K_1$ & 1.3366 & -2.709  & 4.6335   & -6.4748 \\
                       & $K_2$ & 1.7417 & -3.6212 & 5.7109   & -10.468 \\
\enddata
\tablecomments{\footnotesize{}}
\label{tab:K1K2PolynomialCoeff}
\end{deluxetable}

\begin{figure}[ht]
\centering
\includegraphics[trim=0.2cm 4.3cm 0.01cm 5cm , scale=0.55, clip]{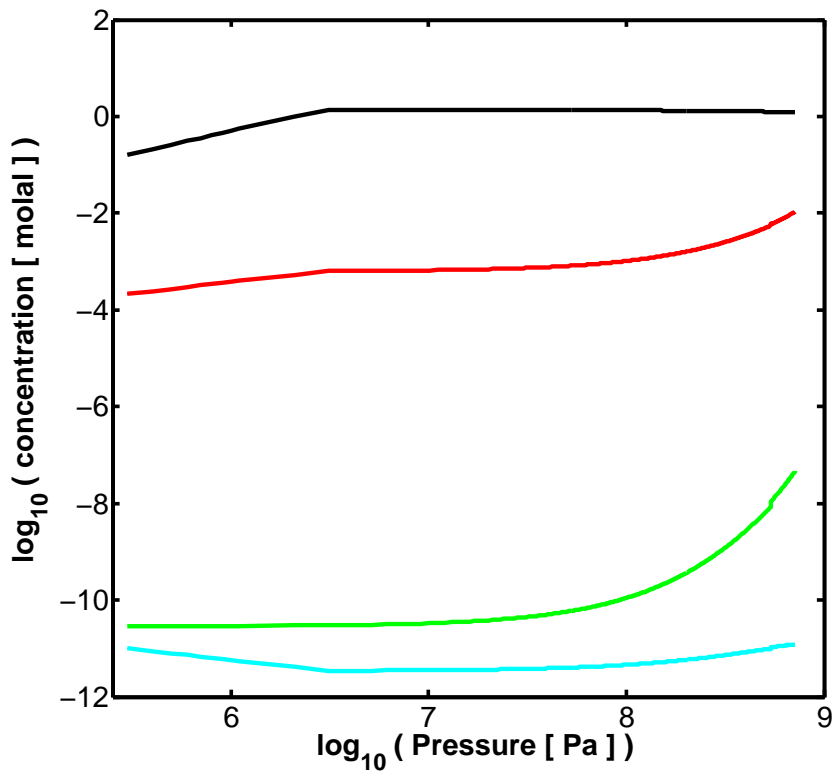}
\caption{\footnotesize{The speciation of the carbonate system as a function of depth in the ocean. Here we consider the ocean thermal profile to be an isotherm, set at $280$\,K. Black curve is the concentration of freely dissolved CO$_2$. Red curve is the concentration of H$^+$ and HCO$_3^{-}$. Green curve is the concentration of CO$_3^{2-}$. Cyan curve is the concentration of OH$^-$. The curves terminate at the pressure representing the bottom of the ocean, for the given isotherm.}}
\label{fig:CarbonateSpeciation}
\end{figure}

In Fig.\ref{fig:CarbonateSpeciation} we give the speciation of the carbonate system as a function of depth, for the $280$\,K isotherm. We find that freely dissolved CO$_2$ is the most abundant form of dissolved inorganic carbon ($DIC$), in water planet oceans, at any depth. This is opposite to the case of Earth, where freely dissolved CO$_2$ is the least abundant $DIC$ component in the ocean \citep{zeebe2001}. 
Also, we find no substantial difference between the concentrations of hydrogen complexes, H$^+$, and bicarbonate (both are represented by the red curve). Carbonate and hydroxide ions are found to exist in the ocean only in trace amounts. We note, that the errors we have suggested above for the values of $K_1$ and $K_2$ do not have any considerable effect on our results.

From Fig.\ref{fig:CarbonateSpeciation} we see that the different chemical species making up the carbonate system have a concentration gradient with depth.
We argue that these concentration gradients with depth are stable against vertical mixing in an ocean world.
First, we wish to point out, that the concentration of freely dissolved CO$_2$, the most abundant component of the $DIC$, is constant to a good approximation throughout the ocean. This is a consequence of the nature of solubility when in equilibrium with the clathrate phase, rather then due to efficient vertical mixing in the ocean \citep{Levi2017}. Thus, the concentration gradients with depth are most pronounced for the chemical species that are least abundant.  
If a parcel of ocean water migrates from the ocean floor upward, this parcel would be supersaturated with carbonate with respect to the shallower depth. The value for $K_2$ decreases with the decreasing pressure. Consequently, carbonate would transform to bicarbonate, and then to freely dissolved CO$_2$. However, this newly formed CO$_2$ is added to an ocean likely saturated with CO$_2$, at equilibrium levels with respect to its clathrate hydrate phase \citep{Levi2017}. Therefore, the additional dissolved CO$_2$ forms grains of clathrate hydrates, that tend to sink due to their high density \citep{Levi2017}. This mechanism keeps the carbonate speciation at every depth at the local equilibrium values. Only a very vigorous vertical mixing of the ocean can counteract this effect and homogenize the ocean. However, the depth scale of the ocean (tens of kilometres) and the lack of external forces capable of driving such a mixing, suggest that homogenization is unlikely \citep[see section $5$ in][]{Levi2017}.     

\begin{figure}[ht]
\centering
\includegraphics[trim=0.2cm 4.3cm 0.01cm 5cm , scale=0.55, clip]{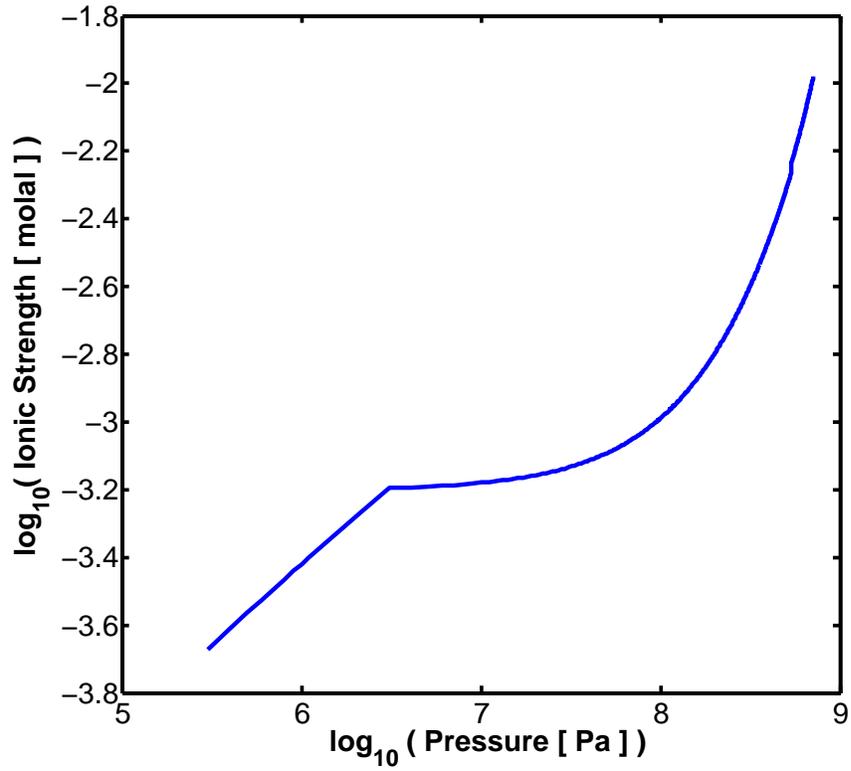}
\caption{\footnotesize{Ionic strength versus depth, assuming for the ocean an isotherm of $280$\,K.}}
\label{fig:IonicStrength}
\end{figure}

We check for the consistency of our results with our initial estimation that the activity coefficients for the different ions can be modelled using the Davies equation. From the variation of the ionic strength with depth, shown in Fig.\ref{fig:IonicStrength}, we see that the ionic strength is well within the regime for which the Davies equation applies.

\begin{figure}[ht]
\centering
\includegraphics[trim=0.2cm 4.3cm 0.2cm 5cm , scale=0.55, clip]{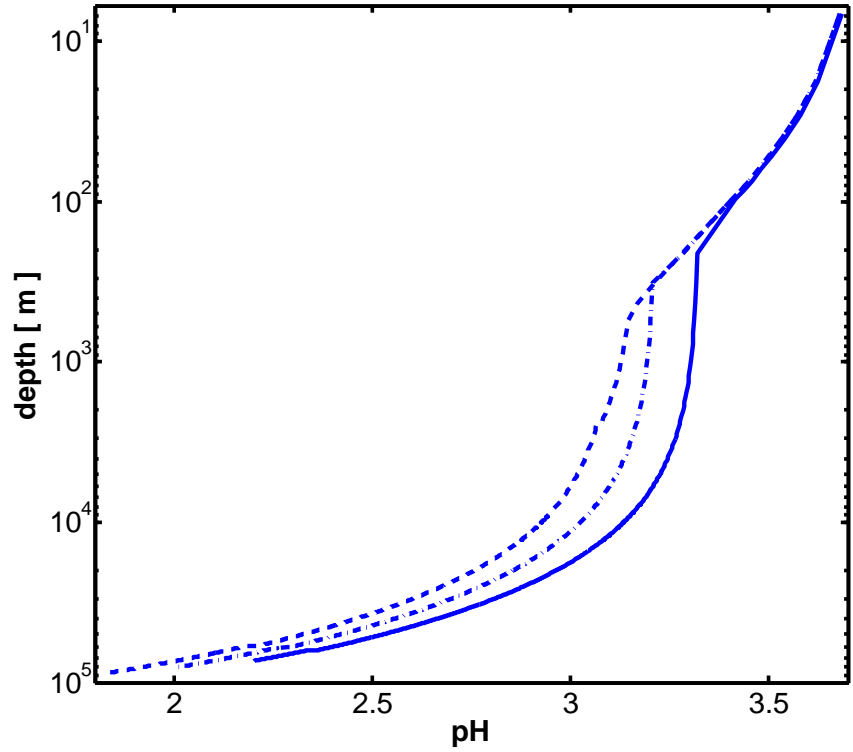}
\caption{\footnotesize{Ocean pH versus depth, assuming the ocean is isothermal. The acidity is increasing with depth in the ocean. Solid curve is for an isotherm of $275$\,K. Dashed-dotted curve is for an isotherm of $280$\,K. Dashed curve is for an isotherm of $285$\,K.}}
\label{fig:OceanpH}
\end{figure}

In Fig. \ref{fig:OceanpH} we plot the ocean pH versus depth (measured positive into the planet). The pH is derived by taking the negative of the common logarithm of the activity for H$^+$. It is clear the ocean is acidic, with an increasing acidity with depth. In case the carbonate system speciation is indeed stably stratified, as discussed above, then so is the pH of the ocean.

\section{DISCUSSION}

In this work we have focused on the binary H$_2$O-NaCl. It is of interest to consider the extension of our conclusions to other salts. 
The efficiency of the desalination pump depends on the value of the solubility of salts in high pressure water ice, at premelting conditions. The higher the solubility is, the larger is the fractionation factor, $f_1$, and the more rapidly the ocean is depleted of its ionic species.

\cite{Klotz2016} reported that an aqueous solution of about $10$\,molal of LiCl and LiBr solidified into ice VII at room temperature. This value for the solubility is much larger than what is found for the case of NaCl in water ice VII, though, this may depend on the particulars of the experiment. 
The solubility of RbI in ice VII is lower than the value reported for NaCl \citep{Frank2006,Journaux2017}. There seems to be some variability in the fractionation factor of salts into high pressure ice, depending on the chemical species considered. We find that the desalination time scale is less than $1$\,Gyr as long as the fractionation factor is more than $0.004$, and less than $2$\,Gyr for a fractionation factor higher than $0.002$. Experimental results indicate that even the relatively low fractionation factor appropriate for ice VI is larger than these values. However, further experiments are needed for various ionic species in order to discover whether for some the fractionation factor is lower than the above values. If that is the case, ocean worlds may be enriched with these species relative to others. Further experiments on more complex systems including not only salts but clathrate hydrate promoters are necessary.    

It is interesting to estimate whether external sources, such as a meteoritic flux, can replenish the ocean salt content and counteract the desalination pump. 
For example, magnesium and sodium sulfates are abundant in carbonaceous chondrites, and highly soluble in water, even at low temperatures \citep{Mason1963,Kargel1991}. Therefore, there is reason to believe that SO$^{2-}_4$ ions may reach our studied oceans with meteorites.
We note that, the radius of SO$^{2-}_4$ is quite large. Thus, it may have a low solubility in high pressure water ice and therefore less effected by the desalination pump. 

There are many bacteria that can multiply in distilled water supplemented with a minimal amount of organic nutrients, with hardly any salt added. Genera of bacteria well known for this are e.g. Hyphomicrobium and Caulobacter \citep[see][for PCa medium with $8\times 10^{-4}$\,molal of MgSO$_4$]{Poindexter2006}. We note that the desalination pump gives an upper bound value of millimolal for the concentration of salt, for planets that are $5$\,Gyr old. Considering an ocean $80$\,km deep, and a planetary radius of $8500$\,km, about $7\times 10^{18}$\,kg of MgSO$_4$ are needed to create a solution of $8\times 10^{-4}$\,molal. Approximately one fifth of the mass of C1 chondrites is salt, dominated by metal sulfates, and epsomite (MgSO$_4\cdot 7$H$_2$O) in particular. This translates to a required mass of C1 chondrites of $7\times 10^{19}$\,kg, if chondritic salt can dissolve entirely in the ocean. In other words, assuming a constant influx of meteorites over $1$\,Gyr, an influx of $80$\,kg\,km$^{-2}$\,yr$^{-1}$ is needed. \cite{Schmitz2001} report on a meteoritic mass flux of about $10^{-3}$\,kg\,km$^{-2}$\,yr$^{-1}$ falling on Earth's surface in the early ordovician. \cite{Bland2003} calculated a present day influx on Earth's surface of $2\times 10^{-6}$\,kg\,km$^{-2}$\,yr$^{-1}$. These values are much lower than what is required. 
Clearly, if the influx of meteoritic material falling on Earth is true for our studied planets as well, then it may not be considered a substantial source for sulfates.

If certain ionic species are found to have a fractionation factor less than what is stated above, then they ought be inefficiently depleted out of the ocean. Hence, chemical reactions involving these ions should be added to the analysis in subsection $4.1$. In addition, their effect on the activity coefficients discussed in subsection $4.1$ will have to be assessed, because they would likely increase the ionic strength of the solution. This would require an estimation of the concentration of such ions that may have accumulated in the ocean during planetary accretion and differentiation. Such a model is beyond the scope of this work.

We find our studied oceans are acidic. Acidophiles grow optimally at pH$<3$, by maintaining pH homeostasis. Because an acidic environment has a high concentration of H$^+$, maintaining homeostasis requires moderating the influx of H$^+$ into the cytoplasm. This may be achieved by concentrating monovalent cations within the cell, often using K$^+$. This creates a potential barrier called the Donnan potential \citep{Craig2007}. A desalination pump, working efficiently in removing ions from the ocean may render this mechanism ineffective, and create a very inhospitable environment.

\begin{figure}[ht]
\centering
\includegraphics[trim=0.2cm 4.3cm 0.2cm 5cm , scale=0.55, clip]{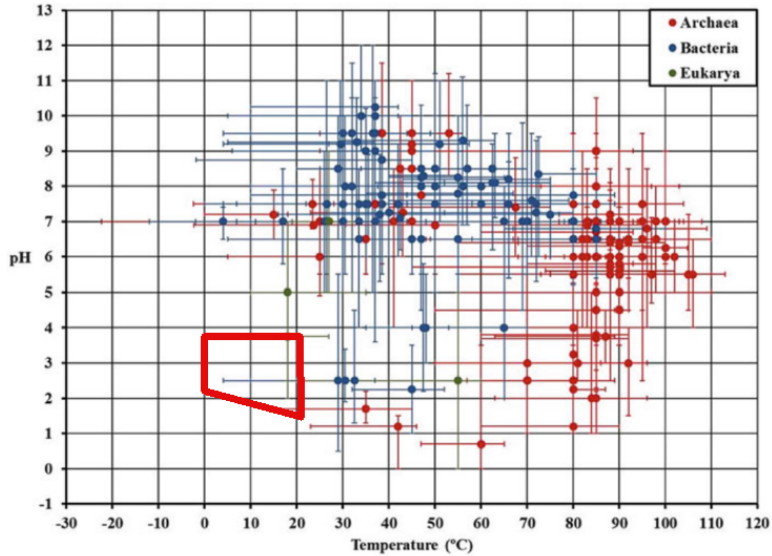}
\caption{\footnotesize{Polyextremophiles in a temperature-pH domain taken from \cite{Polyextremophiles}. The temperature and pH conditions within our studied oceans is demarcated in red. }}
\label{fig:TempVpHdemarkated}
\end{figure}

\cite{Polyextremophiles} tabulated the extremophiles known to date, and plotted them on a temperature-pH diagram (see their figure $5$). In Fig. \ref{fig:TempVpHdemarkated} we demarcate on their diagram (with permission) the temperature and pH conditions likely encountered in our studied exoplanetary oceans. Clearly, the conditions within these oceans are poly-extreme, both low temperature and low pH values. However, a few extremophiles do inhabit our demarcated region. The scarcity of extremophiles in the low temperature and acidic regime may be due to the high stress imposed by this regime on life. However, it may also be the result of the scarcity of such an environment in Earth's geological history \citep{Polyextremophiles}. We note that, in our studied oceans there may be additional stressors, such as: high pressure and low salinity. These are not represented in Fig. \ref{fig:TempVpHdemarkated}. Can polyextremophiles form in our exoplanetary oceans? The unique stressors in our studied oceans may have never existed on Earth. In this case, perhaps, the answer is hidden within the realm of synthetic biology.        

In this work we solved for a $2$M$_\Earth$ planet, however, the desalination pump may prove to be applicable beyond this mass range. Our suggested pump depends on the existence of high pressure ice underlying the ocean, overturn in the water ice mantle, and exceeding a temperature of about $500$\,K in the water ice mantle interior. These requirements are likely met in the interiors of ocean worlds with more than a few percent of their mass in water. However, for more massive planets a likely dense atmosphere of H/He may render internal temperatures too high, not allowing for high pressure water ice polymorphs in the interior. For these planets this work does not apply.

We call for more experiments on high pressure brine and its associated ice at near melting conditions. It is these conditions that likely control the exchange of matter of biological importance between the ocean and the deep water ice mantle, in what may prove to be a very common type of planet.

\section{SUMMARY}

In this work we have focused on ocean worlds, defined here as planets with a global surface ocean, that is separated from any rocky interior by a layer of high-pressure ice polymorphs. We determine the temperature profile and dynamics of this high pressure ice mantle (see section $2$). We find that convection in the water ice mantle can promote resurfacing and downwelling of the ocean floor. This results in the formation of ridges along the ocean floor. Beneath these ridges upwelled high-pressure ice melts, discharging warm water into the ocean at a rate of at least $10^{8}$\,kg\,s$^{-1}$.
In a steady-state, where the mass of the ocean is constant, this rate of melting is balanced by an equal rate of ocean solidification. Therefore, ocean water go through a cycle of melting and solidification.  

In section $3$ we have estimated the concentration of salt dissolved in the ocean, with an emphasis on the H$_2$O-NaCl system. We find that gravity, due to density differences, imposes a upper bound value on the salinity of about $0.5$\,molal of NaCl.   
Low pressure experiments indicate that salt is excluded out of growing ice Ih crystals almost completely \citep[e.g.][]{Lake1970,Worster1997}. This is not the case for high pressure ice. During the melting and solidification cycle of ocean water, hydrated ions from the ocean become incorporated in the newly forming ocean floor, predominantly within an on-melt water ice layer, and at plate bending regions. As this new ocean floor folds into the deep ice mantle, conditions for the exsolution of the salt ions are met, resulting in the formation of grains of salt within the high-pressure ice mantle. These grains are dense and sediment out of the ice mantle. We further show that brine pockets forming on the ice-rock mantle boundary cannot migrate upward to the ocean. This constitutes a pump, which desalinates our studied oceans over time (see illustration in Fig.\ref{fig:DesalinationPump} ). We find the pump has a time scale of the order of tens of millions of years if the ocean floor is made of ice VII, and hundreds of millions of years if the ocean floor is composed of ice VI. Therefore, after a few Gyr our studied oceans may become very poor in salts.

In section $4$ we discuss the difference between open and closed systems. We show that the oceans we study are open systems with regards to CO$_2$. In subsection $4.1$ we introduce the equations governing the pH and carbonate system speciation of the ocean. In subsection $4.2$ we estimate the first and second dissociation constants of carbonic acid for our entire parameter space of interest. We discuss two models for the dissociation constants: the solvent density model, and the revised-HKF model. We show that the solvent density model is appropriate for interpolations between experimental data points, and demonstrate the dangers of applying it as a tool to make extrapolations. For the latter purpose, we show that the revised-HKF model is far more reliable.

In section $5$ we solve for the oceans' pH and carbonate system speciation. We find that freely dissolved CO$_2$ is the dominant component in the dissolved inorganic carbon (DIC). The second most abundant component is bicarbonate, whereas carbonate is present only in trace amounts. The ratios between these three chemical species change somewhat with depth in the ocean. Nonetheless, not enough to change this general picture. We find our oceans to be acidic, with surface water (pH$\approx 3.5$) less acidic than the ocean abyss (pH$\approx 2$).

\section{ACKNOWLEDGEMENTS}

We wish to thank Prof. William B. Durham for a helpful conversation on the rheology of water ice. We also thank our referee for a careful read of the manuscript and helpful suggestions. This work was supported by a grant from the Simons Foundation  (SCOL No.$290360$) to Prof. Dimitar Sasselov.

\section{APPENDIX}

\subsection{APPENDIX A: THE MELTING POINT DEPRESSION}

Here we derive the melting point depression equation we use in this work. On the melting curve of ice VII, the chemical potential of water in ice VII and in the fluid solution (with ionic solutes) is equal:
\begin{equation}\label{equilibriumcond}
\mu_{ice}=\mu^{sol}_{H_2O}
\end{equation}
The chemical potential of water in the fluid solution is related to that of pure water fluid \citep{Denbigh}:
\begin{equation}\label{nonidealsol}
\mu^{sol}_{H_2O}=\mu^{pure}_{H_2O}+kT\ln\left(\gamma_{H_2O}(1-x)\right)
\end{equation} 
where $k$ is Boltzmann's constant, $T$ is the temperature, $\gamma_{H_2O}$ is the activity coefficient of water and $x$ is the mole fraction of salt dissolved in the fluid solution. 

Let us consider a pressure P for which water ice VII in a pure water system melts at a temperature $T_0$, and when in contact with a water-salt solution it melts at a temperature $T$. Because the melting point depression due to the ionic solutes may be large we expand the chemical potential to second order:
\begin{equation}
\mu(T,P)\approx\mu(T_0,P)+\left(\frac{\partial\mu}{\partial T}\right)(T-T_0)+\frac{1}{2}\left(\frac{\partial^2\mu}{\partial T^2}\right)(T-T_0)^2
\end{equation}
where the derivatives are estimated on the melting curve of the pure water system. In terms of entropy the last equation may be rewritten as:
\begin{equation}
\mu(T,P)\approx\mu(T_0,P)-S(T_0,P)(T-T_0)-\frac{1}{2}\left(\frac{\partial S}{\partial T}\right)(T-T_0)^2
\end{equation}
With the aid of the last expression we may write:
\begin{equation}
\mu_{ice}-\mu^{pure}_{H_2O}\approx S_f(T-T_0)+\frac{1}{2}\left(\frac{\partial S_f}{\partial T}\right)(T-T_0)^2
\end{equation}
Here $S_f$ is the entropy of fusion in the pure water system. Inserting the last relation into Eqs.($\ref{equilibriumcond}$) and ($\ref{nonidealsol}$) and rearranging yields:
\begin{equation}
-S_f\Delta{T}+\frac{1}{2}\left(\frac{\partial S_f}{\partial T}\right)(\Delta{T})^2=k(T_0-\Delta{T})\ln\left(\gamma_{H_2O}(1-x)\right)
\end{equation}
where we have defined the melting point depression $\Delta{T}\equiv T_0-T>0$.

\subsection{APPENDIX B: SALINITY PROFILE IN AN UNMIXED OCEAN}

Here we estimate how the salinity changes with depth in an unmixed deep ocean. Therefore, we assume an ocean in diffusive equilibrium. The chemical potential of the solvent (water) may be written as:
\begin{equation}
\mu_{H_2O}=\mu^{pure}_{H_2O}+kT\ln\left(\gamma_{H_2O}X_{H_2O}\right)+m_wgz
\end{equation}
where $\mu^{pure}_{H_2O}$ is the chemical potential of pure water, $k$ is Boltzmann's constant, $T$ is the temperature, $\gamma_{H_2O}$ is the activity coefficient of water and $X_{H_2O}$ is the mole fraction of water in the fluid solution. The mass of a water molecule is $m_w$, the gravitational acceleration is $g$, and $z$ is the depth in the ocean. 

We write for the solute (NaCl), the following:
\begin{equation}
\mu_{NaCl}=\mu^*_{NaCl}+kT\ln\left(\gamma_{NaCl}X_{NaCl}\right)+\bar{m}_igz
\end{equation}
Here $\mu^*_{NaCl}$ is the chemical potential of the solute at infinite dilution. The activity coefficient and mole fraction of the salt are $\gamma_{NaCl}$ and $X_{NaCl}$, respectively. The average mass of the ions composing the salt is $\bar{m}_i$.  

In a state of diffusive equilibrium the chemical potential of the solvent is equal throughout the ocean. The same is true for the solute. Thus, differentiation with respect to depth yields:
$$
\frac{d}{dz}\mu^{pure}_{H_2O}+kT\frac{1}{X_{H_2O}}\frac{dX_{H_2O}}{dz}+m_wg = 0
$$
\begin{equation}
\frac{d}{dz}\mu^*_{NaCl}+kT\frac{1}{X_{NaCl}}\frac{dX_{NaCl}}{dz}+\bar{m}_ig = 0
\end{equation}
For simplicity we have assumed that the activity coefficients are weakly dependent on the depth in the ocean. We use the relation $X_{H_2O}+X_{NaCl}=1$ and subtract between the last two equations, giving,
\begin{equation}
0=\frac{d}{dz}\left(\mu^{pure}_{H_2O}-\mu^*_{NaCl}\right)-kT\left(\frac{1}{1-X_{NaCl}}+\frac{1}{X_{NaCl}}\right)\frac{dX_{NaCl}}{dz}+g(m_w-\bar{m}_i)
\end{equation}
With the aid of the hydrostatic relation we can write:
\begin{equation}
0=-\rho_{oc} g\frac{d}{dP}\left(\mu^{pure}_{H_2O}-\mu^*_{NaCl}\right)-kT\left(\frac{1}{1-X_{NaCl}}+\frac{1}{X_{NaCl}}\right)\frac{dX_{NaCl}}{dz}+g(m_w-\bar{m}_i)
\end{equation}   
Here, $\rho_{oc}$ is the density of the ocean. However, the pressure derivative of the chemical potential is a volume, yielding the following relation:
\begin{equation}
0=\rho_{oc} g\left(v^{pure}_{H_2O}-v^*_{NaCl}\right)+kT\left(\frac{1}{1-X_{NaCl}}+\frac{1}{X_{NaCl}}\right)\frac{dX_{NaCl}}{dz}+g(\bar{m}_i-m_w)
\end{equation}  
where $v^{pure}_{H_2O}$ is the volume per water molecule in a pure water system, and $v^*_{NaCl}$ is an average volume occupied by the hydrated ions of salt at the limit of infinite dilution.

We consider the bottom of the ocean as a reference level, where the depth is $z_0$ and the salt abundance is $X^0_{NaCl}$. Integrating over the last relation gives:
\begin{equation}
\frac{X_{NaCl}}{1-X_{NaCl}}=\frac{X^0_{NaCl}}{1-X^0_{NaCl}}\exp\left(-\frac{g(\bar{m}_i-m_w)}{kT}(z-z_0)\right)\exp\left(-\frac{g}{kT}\int^z_{z_0}\rho_{oc}\left[v^{pure}_{H_2O}-v^*_{NaCl}\right]dz' \right)
\end{equation}
In terms of the molality of the salt, $m_{NaCl}$, the last equation reads:
\begin{equation}\label{molalityprofile}
m_{NaCl}=m^0_{NaCl}\exp\left(-\frac{g(\bar{m}_i-m_w)}{kT}(z-z_0)\right)\exp\left(-\frac{g}{kT}\int^z_{z_0}\rho_{oc}\left[v^{pure}_{H_2O}-v^*_{NaCl}\right]dz' \right)
\end{equation} 
The integrand in the last term of Eq.($\ref{molalityprofile}$) does not vary considerably with depth in the ocean, for our pressure range of interest (up to $1$\,GPa). This is evident from the radial distribution functions, giving the distances between O-O, Na-O, and Cl-O \citep{Jancso1984}.
However, conservatively, below we shall adopt values that maximize the integrand, thus, overestimating its effect in moderating the salinity gradient in the ocean. This then yields a simpler relation:
$$
m_{NaCl}(z)=m^0_{NaCl}e^{-\frac{z-z_0}{H_s}}
$$
\begin{equation}
\frac{1}{H_s}\equiv\frac{g}{kT}\left\lbrace\bar{m}_i-m_w+\rho_{oc}\left[v^{pure}_{H_2O}-v^*_{NaCl}\right]\right\rbrace
\end{equation}
where $H_s$ is a salinity scale height.

Therefore, the average concentration of salt in the ocean is:
\begin{equation}
\left\langle m_{NaCl} \right\rangle =\frac{1}{R_p-z_0}\int^{R_p}_{z_0}m_{NaCl}(z')dz' = m^0_{NaCl}\frac{H_s}{H_{oc}}\left[1-e^{-H_{oc}/H_s}\right]
\end{equation}
where $R_p$ is the radius of the planet and $H_{oc}$ is the depth of the ocean. For a $2$M$_\Earth$ planet assuming $50$\% of its mass is H$_2$O we take $g=890$\,cm\,s$^{-2}$ \citep{Levi2014}. For the average ionic mass, for the case of NaCl, we have $\bar{m}_i=4.85\times 10^{-23}$\,g. For the density of pure water we assume  the maximal value throughout the ocean, $\rho_{oc}=1.24$\,g\,cm$^{-3}$ \citep[see][for a pressure of $1$\,GPa]{wagner02}. In order to compare volumes in a consistent way we consider the hard sphere model and use data for the first peak from radial distribution functions. For the O-O distance we adopt a value of $2.8$\AA \citep{Katayama2010}. \cite{Mancinelli2007} report values of $2.34$\AA,  and $3.16$\AA, for the Na-O and Cl-O peaks, respectively. However, \cite{Lvov1990} suggest a much larger value of $3.26$\AA, for the hard sphere diameter, for deriving the volume at infinite dilution. Using the value from \cite{Lvov1990} gives $H_s=42$\,km, whereas, adopting the data form \cite{Mancinelli2007} yields $H_s=23$\,km. These correspond to enrichments by a factor of $2.6$ and $4.3$ in the salinity at the bottom of the ocean relative to the ocean average, for $H_{oc}=100$\,km.

\bibliographystyle{aasjournal}
\bibliography{amitmemo} 

\end{document}